\documentclass[11pt]{article}%
\usepackage{amsfonts}
\usepackage{amssymb}
\usepackage{graphicx}
\usepackage{setspace}
\usepackage[round]{natbib}
\usepackage{amsmath}%
\usepackage{amsthm}%
\usepackage{url}
\usepackage{palatino}
\usepackage{paralist}
\usepackage[top=8pc,bottom=8pc,left=8pc,right=8pc]{geometry}

\newtheorem{theorem}{Theorem}

\newtheorem{example}{Example}

\newtheorem{lemma}[theorem]{Lemma}

\def\E{\qopname\relax o{E}}

\newcommand{\Ga}{ \mathcal{G} }
\newcommand{\N}{ \mathcal{N} }
\newcommand{\DE}{ \mathcal{DE} }
\newcommand{\Ex}{ \mathcal{E} }
\newcommand{\prox}{ \mathop{\mathrm{prox}} }
\newcommand{\R}{\mathcal{R}}

\newcommand{\sgn}{\mathop{\mathrm{sgn}}}
\newcommand{\dd}{\mathrm{d}}

\newcommand{\enorm}[1]{\Vert #1 \Vert_2}

\title{Mixtures, envelopes, and hierarchical duality}
\author{Nicholas G. Polson\\
\textit{Booth School of Business}\\
\textit{University of Chicago}\\
\\
James G. Scott\\
\textit{McCombs School of Business}\\
\textit{University of Texas at Austin}
\footnote{Polson is Professor of Econometrics and Statistics
at the Chicago Booth School of Business. email: ngp@chicagobooth.edu. Scott
is Assistant Professor of Statistics at the McCombs School of Business, University of Texas at Austin.
email: James.Scott@mccombs.utexas.edu.  Scott's work has been partially supported by a CAREER grant from the U.S.~National Science Foundation (DMS-1255187). }
}
\date{February 2015}

\begin{document}

\maketitle
\begin{abstract}
\noindent We develop a connection between mixture and envelope representations of objective functions that arise frequently in statistics.  We refer to this connection using the term ``hierarchical duality.''  Our results suggest an interesting and previously under-exploited relationship between marginalization and profiling, or equivalently between the Fenchel--Moreau theorem for convex functions and the Bernstein--Widder theorem for Laplace transforms.  We give several different sets of conditions under which such a duality result obtains.  We then extend existing work on envelope representations in several ways, including novel generalizations to variance-mean models and to multivariate Gaussian location models. This turns out to provide an elegant missing-data interpretation of the proximal gradient method, a widely used algorithm in machine learning.  We show several statistical applications in which the proposed framework leads to easily implemented algorithms, including a robust version of the fused lasso, nonlinear quantile regression via trend filtering, and the binomial fused double Pareto model.  Code for the examples is available on GitHub at \url{https://github.com/jgscott/hierduals}.

\vspace{0.5pc}

\noindent {\bf Keywords:} Bayesian inference, convex duality, envelopes, MAP estimation, Gaussian mixtures, penalized likelihood, variational methods
\end{abstract}

\newpage

\section{Introduction}

\subsection{Marginalization versus profiling in hierarchical models}

A large number of statistical problems can be expressed in the form
\begin{equation}
\label{eqn:canonicalform}
\begin{aligned}
& \underset{x \in \R^d}{\text{minimize}}
& & l(x) + \phi(x) \, .
\end{aligned}
\end{equation}
Perhaps the most common example arises in estimating a generalized linear model, where $l(x)$ is the negative log likelihood and $\phi(x)$ is a penalty function that regularizes the estimate.   From the Bayesian perspective, the solution to this problem may be interpreted as a maximum \textit{a posteriori} (MAP) estimate in the hierarchical model
\begin{equation}
\label{eqn:bayesversion}
p(y \mid x) \propto \exp\{-l(x)\} \; , \quad p(x) \propto \exp\{ -\phi(x) \} \, .
\end{equation}
Another case of (\ref{eqn:canonicalform}) arises in decision problems where options are compared based on expected loss, and $l(x)$ and $\phi(x)$ represent conceptually distinct contributions to the loss function: $l(x)$ is tied to the data and $\phi(x)$ to the cost associated of the decision.  Many Bayesian testing and model-selection problems can be phrased in just this form \citep{scottberger06,muller:etal:2006,hahn:carvalho:2013}.
 
This paper is about the use of auxiliary variable schemes for representing probability models such as (\ref{eqn:bayesversion}) in analytically convenient forms.  Our examples focus on regression and smoothing problems.  But because (\ref{eqn:canonicalform}) and (\ref{eqn:bayesversion}) have the same optimal points, such schemes can be useful for any statistical optimization problem of the form (\ref{eqn:canonicalform}). 

Specifically, we study the connection between mixture and envelope representations of statistical objective functions.  A mixture is the marginal of a higher-dimensional joint distribution: $p(x) = \int_\Lambda p(x, \lambda) \ d \lambda$.  An envelope (or variational representation) is the pointwise supremum of a higher-dimensional joint distribution: $p(x) = \sup_\lambda \{ p(x ,\lambda) \}$.    These two representations correspond to the statistical operations of marginalizing and profiling out an auxiliary variable $\lambda$, respectively.  Each approach has a long statistical tradition in its own right, reflecting a particular school of thought about how to handle nuisance parameters: marginalization is stereotypically Bayesian, while profiling is stereotypically frequentist, seemingly without a natural Bayesian interpretation.

In the context of auxiliary-variable representations of (\ref{eqn:bayesversion}), however, marginalizing and profiling are merely two ways of defining one family of probability distributions in terms of another.  In fact, we will show that there are many cases in which these two operations are dual to each other, in the sense that profiling under one model corresponds to marginalizing under a different model.  This establishes a formal Bayesian interpretation of profiling in several important special cases.

This paper makes the following specific contributions.  First, we give several different sets of conditions under which marginalization and profiling are dual to one another.  These sets of conditions correspond to different forms for the joint model $p(x, \lambda)$.  While they are not exhaustive, they still encompass a wide variety of practical problems.  We consider conditionally exponential models (Section 2) and conditionally normal models (Section 3).  In this context, one of our primary goals is to encourage Bayesians to view the EM algorithm as just one of a broader family of optimization procedures with elegant missing-data interpretations.

Second, we give conditions under which $p(x)$ may be represented as a variance-mean envelope of Gaussian distributions.  This extends some of our own work on the use of variance and variance-mean mixtures in Bayesian computation \citep{Polson:Scott:2010b,Polson:Scott:2011a}.  It also generalizes work by \citet{geman:reynolds:1992} and \citet{geman:yang:1995} on the class of half-quadratic penalties by connecting it with more recent work on penalized likelihood \citep[e.g.][]{taddy:2010,strawderman:etal:2013}.

Third, we propose a multivariate generalization of envelope representations based on Gaussian location models.  This provides an interesting statistical interpretation of the proximal gradient method, a widely studied algorithm in the literature on signal processing.

Finally, we demonstrate some interesting statistical applications of the approach.  These applications highlight the strength of our framework: the way it allows practitioners to ``plug and play'' by mixing likelihoods and penalties with little analytical work, while staying in an algorithmic framework whose building blocks are familiar and efficient (e.g.~weighted least squares and soft thresholding).  We show this on three examples: an outlier-robust fused lasso; nonlinear quantile regression via trend filtering; and the fused double-Pareto model for nearly unbiased spatial smoothing of binomial outcomes.

One thing we do not do is to study the frequentist properties of estimators.  Rather, we focus on the representation of objective functions in terms of notionally missing data, and on the algorithmic consequences of such representations.  Good recent examples of work on the frequentist properties of Bayesian shrinkage rules include \citet{bogdan:ghosh:2008b},  \cite{bhattacharya:etal:2012}, and \citet{datta:ghosh:2013}, and we encourage the interested reader to consult these papers.

\subsection{Related work}

Our paper builds upon five papers in particular: those by \citet{geman:reynolds:1992}, \citet{geman:yang:1995}, \citet{Polson:Scott:2011a}, \citet{taddy:2010}, and \citet{strawderman:etal:2013}.  These works discuss the fundamental problem of representing probability distributions in algorithmically useful ways by introducing latent variables.  Our goal is to provide a more general theory that unites these various representations under the framework of hierarchical models, and to demonstrate the statistical applications of this framework.

There has also been recent interest in the Bayesian literature in representing likelihoods and pseudo-likelihoods using mixtures \citep{li:xi:lin:2010,polson:stevescott:2011,gramacy:polson:2012,polson:scott:windle:2012a}, and our paper sits firmly in this line of work as well.  Algorithms that exploit mixture representations of probability densities are very common in Bayesian inference.  For example, it is typical to express Bayesian versions of penalized-likelihood estimators as Gaussian scale mixtures.  See, for example, the papers on the Bayesian lasso estimator \citep{park:casella:2008,hans:2008}; the bridge estimator \citep{Polson:Scott:2011d}; the relevance vector machine of \citet{tipping:2001}; the normal/Jeffreys prior of \citet{figueiredo:2003} and \citet{bae:mallick:2004}; the normal/exponential-gamma model of \citet{griffin:brown:2005}; the normal/gamma and normal/inverse-Gaussian models \citep{caron:doucet:2008, griffin:brown:2010}; the horseshoe prior of \citet{Carvalho:Polson:Scott:2008a}; the double-Pareto model of \citet{dunson:armagan:lee:2010}; and the Bayesian elastic net \citep{hans:2011}.  Envelope representations are also commonly used in variational-Bayes approximations of posterior distributions \citep[e.g.][]{jaakkola:jordan:2000,armagan:2009}.

\subsection{Preliminaries}

We begin by establishing some definitions, notation, and important facts.  We use $y$ to denotate an $n$-vector of outcomes and $A$ a fixed $n \times d$ matrix whose rows $a_i^T$ are the design points or features.  All vectors are column vectors.  Observations are indexed by $i \in \{1, \ldots, n\}$, parameters by $j \in \{1, \ldots, d\}$ and iterations in an algorithm by $t \in \mathbb{N}$.

All functions in this paper are assumed to be lower semi-continuous.  We also use the following conventions: $\sgn(x)$ is the algebraic sign of $x$; $x_+ = \max(x, 0)$; $\iota_C(x)$ is the set indicator function taking the value $0$ if $x \in C$, and $\infty$ if $x \notin C$; $\mathcal{R}^+ = [0, \infty)$, $\mathcal{R}^{++} = (0, \infty)$, and $\overline{\mathcal{R}}$ is the extended real line $\mathcal{R} \cup \{-\infty, \infty\}$.  We use $\Vert v \Vert_a$ to denote the $\ell^a$ norm of a vector,
$$
\Vert v \Vert_a = (|v_1|^a + \cdots + |v_d|^a)^{1/a} \, ,
$$
and $x^T y$ for the Euclidean inner product between two vectors. The soft thresholding operator is denoted by
\begin{equation}
\label{eqn:soft.threshold}
S(y; \lambda) = \arg \min_{x} \left\{ \frac{1}{2}(y-x)^2 + \lambda |x| \right\} = \left[ y -  \lambda \sgn(y) \right]_+ \, .
\end{equation}

We use $\N(x \mid \mu, \sigma^2)$ to denote the density function, evaluated at $x$, of the normal distribution with mean $\mu$ and variance $\sigma^2$.  Similary, $\Ga(x\mid r,s)$ is the gamma distribution with shape $r$ and rate $s$, $\Ex(x \mid r)$ the exponential distribution with rate $r$, and $\DE(x \mid m, s)$ the double-exponential or Laplace distribution with center $m$ and scale $s$.  Where it is clear from context, we will also use $\N(x \mid \mu, \Sigma)$ to denote the density of the multivariate normal distribution with mean vector $\mu$ and covariance matrix $\Sigma$.


Our discussion of envelope representations requires several concepts from convex analysis.  First, the subdifferential of a convex function $f(x)$ at a point $x_0$ is the set
$$
\partial f(x_0) = \{  \xi : f(x) \geq f(x_0) + \xi^T(x-x_0) \} \, .
$$
If $f$ is differentiable at $x_0$, then $\partial f(x_0)$ is the singleton set containing the ordinary gradient from differential calculus: $\partial f(x_0) = \{\nabla f(x_0) \}$.  By analogy, for a concave function $f$ we define the superdifferential as the set $\{  \xi : f(x) \leq f(x_0) + \xi^T(x-x_0) \}$.  Where the context makes its meaning clear, we will use the notation $\partial f(x)$ to denote both the subdifferential of a convex function and the superdifferential of a concave function.

Another important notion is the \textit{convex conjugate} of a function $f(x)$, defined as $f^{\star}(\lambda) = \sup_x \{ \lambda^T x - f(x) \}$.  As $f^{\star}$ is the pointwise supremum of a family of affine (and therefore convex) functions, it is convex even when $f(x)$ is not.  The following result is called the Fenchel--Moreau theorem.  It is a well-known fact about convex conjugates of closed, proper convex functions \citep[e.g.][\S 3.3.2]{boyd:vandenberghe:2004}.

\begin{lemma}[Fenchel duals]
\label{fact:fenchel}
\begin{enumerate}[(A)]
\item Let $f(x): \R^d \to \overline{\mathcal{R}}$ be a closed convex function.  Then there exists a convex function $f^{\star}(\lambda)$ such that the following dual relationship holds:
\begin{eqnarray*}
f(x) &=& \sup_{\lambda} \{ \lambda^T x - f^{\star}(\lambda) \} \\
f^{\star}(\lambda) &=& \sup_{x} \{ \lambda^T x - f(x) \} \, .
\end{eqnarray*}
\item If $f(x)$ is instead a concave function, (A) holds with $\sup$ replaced by $\inf$ in both equations.
\item Any maximizing value of $\lambda$ in Part (A) satisfies
$$
\hat{\lambda} \in \hat{\Lambda}(x) \iff \hat \lambda \in \partial f(x) \, ,
$$
where
$$
\hat{\Lambda}(x) = \left\{ \hat{\lambda} : \hat{\lambda}^T x - f^{\star}(\hat{\lambda}) = \sup_{\lambda} [ \lambda^T x - f^{\star}(\lambda) ]  \right\} \, .
$$
\end{enumerate}
\end{lemma}

Part (B) follows by applying claim (A) to the convex function $-f(x)$, and appealing to the fact that $\sup_A g(x) = -\inf_A[-g(x)]$.  Part (C) follows directly from the Fenchel--Young inequality; see, for example, Proposition 11.3 of \citet{rockafellar:wets:1998}.  As a corollary of Part C, if $f(x)$ is differentiable, the maximizing value of $\lambda$ in the first equation is $\hat{\lambda}(x) = \nabla f(x)$.

One function $g(x)$ is said to majorize another function $f(x)$ at $x_0$ if $g(x_0) = f(x_0)$ and $g(x) \geq f(x)$ for all $x \neq x_0$.  If the same relation holds with the inequality sign flipped, then $g(x)$ is instead said to minorize $f(x)$.  A function $f(x)$ is completely monotone on $A \subset \mathcal{R}$ if its derivatives alternate in sign: $(-1)^k f^{(t)}(x) \geq 0$ for all $k = 0, 1, 2, \ldots$ and for all $x \in A$.  A completely monotone function is therefore nonnegative, nonincreasing, convex, and so forth.

Finally, for any function $f(x)$, the Moreau envelope $E_\gamma f(x)$ and proximal mapping $\prox_{\gamma} f(x)$ for parameter $\gamma > 0$ are defined as
\begin{eqnarray}
E_{\gamma f} (x) &=& \inf_{z } \left\{f(z) + \frac{1}{2\gamma} \enorm{z - x}^2  \right\}  \leq f(x) \\
\prox_{\gamma f} (x) &=& \arg \min_{z } \left\{  f(z)+ \frac{1}{2\gamma} \enorm{z - x}^2  \right\} \, .
\end{eqnarray}
Intuitively, the Moreau envelope is a regularized version of $f$.  It approximates $f$ from below, and has the same set of minimizing values as $f$ \citep[Chapter 1G]{rockafellar:wets:1998}.  The proximal mapping returns the value that solves the minimization problem defined by the Moreau envelope.  It balances two goals: minimizing $f$, and staying near $x$.  The proximal operator generalizes the notion of Euclidean projection onto a convex set: if $f(x) = \iota_C(x)$, then $\prox_f(x) = \arg \min_{z \in C}  \Vert x-z \Vert_2^2$.  Many intermediate steps in the algorithms we discuss below have compact expressions in terms of proximal operators of known functions.

\section{Hierarchical duality: the exponential case}
\label{section:exponential}
\subsection{Envelope representations for concave penalties}

Starting with conditionally exponential models, we develop a connection between mixture and envelope representations, or equivalently between marginalization and profiling.  To keep the focus on the essential ideas, we assume a scalar Gaussian noise model $(y \mid x) \sim \N(y \mid x, 1)$, so that the MAP inference problem is
\begin{equation}
\label{eqn:penalizedLS}
\begin{aligned}
& \underset{x \in \mathcal{R}}{\text{minimize}}
& &  \left\{ \frac{1}{2}(y-x)^2 + \phi(x) \right\} \, .
\end{aligned}
\end{equation}

Suppose that $\phi(x)$ is a symmetric nonnegative function and that is concave and nondecreasing on $\mathcal{R}^+$.  Without loss of generality we consider only nonnegative arguments for $\phi(x)$ and thus write $\phi(|x|)$ below.  Any such $\phi$ may be represented in terms of its concave dual $\phi^\star$:
\begin{equation}
\label{eqn:concavevariational}
\begin{aligned}
\phi(|x|) &= \inf_{\lambda \geq 0} \{ \lambda |x| - \phi^\star(\lambda) \} \\
\phi^{\star}(\lambda) &= \inf_{x \geq 0} \{ \lambda x - \phi(|x|) \} \, .
\end{aligned}
\end{equation}
The domain restriction for $\lambda$ is inherited from the fact that $\phi(x)$ is symmetric, nonnegative, and nondecreasing on $\mathcal{R}^+$, which together imply that $\phi^\star(\lambda) = - \infty$ whenever $\lambda < 0$.

The envelope representation suggests a simple iterative algorithm for solving (\ref{eqn:penalizedLS}), and therefore for evaluating the proximal operator $\prox_\phi(y)$. To see this, use (\ref{eqn:concavevariational}) to rewrite the original problem (\ref{eqn:penalizedLS}) as
\begin{equation}
\label{eqn:simplel1profile}
\begin{aligned}
& \underset{x, \lambda }{\text{minimize}}
& &   \left\{ \frac{1}{2}(y-x)^2  +  \lambda |x| - \phi^{\star}(\lambda) \right\} \, .
\end{aligned}
\end{equation}
A local minimum may now be found by iteratively minimizing over $x$ and $\lambda$.  The partial minimization step in $x$ (holding $\lambda$ fixed) is equivalent to solving an $\ell^1$-penalized least-squares fit. In this simple case,
$$
x^{(t+1)} = \arg \min_{x} \left\{ \frac{1}{2}(y-x)^2  +  \lambda^{(t)} |x| \right\} = S(y; \lambda^{(t)})
$$
The partial minimization step in $\lambda$ for fixed $x$ is given by Part C of Lemma \ref{fact:fenchel} as
$$
\lambda^{(t+1)} =  \phi' \big( |x^{(t+1)}| \big) \, .
$$
If $\phi$ is not differentiable at $x$, then we simply replace $\phi'(|x|)$ in the above expression with any element of the superdifferential $\partial \phi(|x|)$.

Notice that with $\lambda^{(t)}$ implicitly defined as a function of $x^{(t)}$, the concavity of $\phi(x)$ means that, for all $x$,
$$
\frac{1}{2}(y-x)^2 + \phi(|x|) \leq \frac{1}{2}(y-x)^2 + \lambda^{(t)} |x| - \phi^{\star}(\lambda^{(t)}) \, ,
$$
with equality achieved at $x = x^{(t)}$.  Thus the calculation of $\lambda^{(t)}$ is the majorization step in a majorization/minimization (MM) algorithm \citep[see, e.g.][]{zhou:lange:suchard:2010}.  This also highlights the connection with the LLA algorithm of \citet{zou:li:2008}.  Consider the variational representation of $\phi(x_0)$, with $x_0$ fixed. Because the value of $\lambda$ that attains the minimum in Equation (\ref{eqn:concavevariational}) is precisely $\phi'(x_0)$, the envelope representation defines a locally linear approximation to $\phi(x)$ at $x_0$ (hence LLA).  

We prefer the interpretation in terms of proximal operators, which has a missing-data interpretation similar to that of the EM algorithm \citep{dempster:laird:rubin:1977}.  Specifically, we have written the prior for $x$ as the envelope of a joint prior distribution in $x$ and $\lambda$, where $p(x \mid \lambda)$ is double exponential.  The prior for $x$ is recovered by treating $\lambda$ as a nuisance parameter and profiling it out:
$$
p(x) \propto e^{-\phi(|x|)} \propto \sup_{\lambda \geq 0} \left\{ e^{- \lambda |x| + \phi^\star(\lambda)} \right\} \propto \sup_{\lambda \geq 0} \left\{ \DE(x \mid 0, \lambda^{-1}) \ p(\lambda) \right\} \, ,
$$
where $p(\lambda) \propto \lambda^{-1} e^{\phi^{\star}(\lambda)}$.  We refer to priors that can be represented this way as envelopes of double exponentials.  

In problems with more complicated likelihoods, the $x$ update will still involve solving an $\ell^1$-penalized problem, for which many efficient algorithms are available \citep[e.g.][]{efron:LARS:2004,boyd:l1:2007,fried:hastie:tibs:2010}. An important caveat is that this iterative scheme can only be expected to converge to global minimum if the original problem is convex.  Only special circumstances will guarantee global convergence in the case of a nonconvex likelihood or penalty \citep{mazumder:friedman:hastie:2009}.

\subsection{Examples}

Several special cases of this envelope representation have been studied in the literature on sparse regression.  We give two examples.

\begin{example}[log penalty or double-Pareto]
\label{example:doublepareto}
\normalfont
Consider the problem
$$
\begin{aligned}
& \underset{x \in \mathcal{R}}{\text{minimize}}
& & \left\{ \frac{1}{2}(y-x)^2 + \gamma \log (1+ |x |/a)  \right\} \, .
\end{aligned}
$$
The penalty function $\phi(|x|; \gamma, a) = \gamma \log (1 + |x|/a)$ was referred to as the double-Pareto penalty by \citet{dunson:armagan:lee:2010}.  It is concave and nondecreasing on $\mathcal{R}^+$, and has dual
\begin{eqnarray*}
\phi^{\star}(\lambda) &=& \inf_{x \geq 0} \{ \lambda |x| - \gamma \log (1 + |x|/a) \} \\
&=& \gamma \log \lambda - \lambda a + C \, ,
\end{eqnarray*}
where $C$ is a constant not involving $\lambda$.  Thus we may express the original problem as
$$
\begin{aligned}
& \underset{x \in \mathcal{R}, \lambda \geq 0}{\text{minimize}}
& & \left\{ \frac{1}{2}(y-x)^2 + \lambda |x| + \lambda a - \gamma \log \lambda \right\} \, .
\end{aligned}
$$

This has a quasi-Bayesian interpretation: the objective function is the joint posterior density in $(x, \lambda)$ arising from the hierarchical model
\begin{equation}
\label{eqn:jointgammalasso}
(y \mid x) \sim \N(y \mid x, 1) \; , \quad (x \mid \lambda) \sim \DE(x \mid 0, \lambda^{-1}) \; , \quad \lambda \sim \Ga(\lambda \mid \gamma, a) \, .
\end{equation}
The solution to the original problem is the $x$ ordinate of the solution of (\ref{eqn:jointgammalasso}):
$$
\arg_{x} \max_{x, \lambda} \left\{ e^{-\frac{1}{2}(y-x)^2} \lambda e^{-\lambda |x|} \lambda^{\gamma-1} e^{-a\lambda} \right\} = 
\arg_{x} \min_{x, \lambda} \left\{ \frac{1}{2} (y - x)^2 + \lambda |x|  - a \lambda + \gamma \log \lambda \right\} \, .
$$
This justifies the alternative name ``gamma-lasso'' adopted by \citet{taddy:2010}.

\end{example}

\begin{example}[Minimax concave penalty]
\normalfont
Consider problem (\ref{eqn:penalizedLS}) where $\phi(x; \gamma, a)$ is the minimax concave penalty (MCP) function, defined by \citet{zhang:2010} as
$$
\phi(x; \gamma, a) = \gamma \int_0^x \left(1 - \frac{t}{a\gamma}\right)_+ dt = 
\left\{ 
\begin{array}{ll}
\lambda x - \frac{x^2}{2a} & \mbox{if } \; x < a\gamma \, , \\
\frac{a\gamma^2}{2} & \mbox{if } \; x \geq a\gamma \, .
\end{array}
 \right.
$$
This is concave and nondecreasing on $\mathcal{R}^+$, and its dual is easily computed:
$$
\phi^{\star}(\lambda; \gamma, a) = \inf_{x \geq 0} \left\{\lambda x - \phi(x; \gamma, a)  \right\} = -\frac{a}{2} (\lambda - \gamma)^2 \mathbb{I}_{\lambda \leq a\gamma} \, .
$$
This leads to
$$
\phi(|x|; \gamma, a) = \inf_{\lambda \geq 0} \left\{ \lambda |x| + \frac{a}{2}(\lambda - \gamma)^2 \mathbb{I}_{\lambda \leq a\gamma} \right\} \, .
$$
\citet{zhang:etal:2013} use this representation to derive an augmented-Lagrangian method for computing the estimator, while \citet{strawderman:etal:2013} derive this same representation via a different argument.

As with the double-Pareto model, the MCP estimate has a quasi-Bayesian interpretation as the joint MAP estimate in $(x, \lambda)$ under a hierarchical model:
$$
(y \mid x) \sim \N(y \mid x, 1) \; , \quad (x \mid \lambda) \sim \DE(x \mid 0, \lambda^{-1}) \; , \quad p(\lambda) \propto \lambda^{-1} \N(\lambda \mid \gamma, a^{-1}) \mathbb{I}_{a\gamma \geq \lambda \geq 0} \, .
$$
\citet{strawderman:etal:2013} refer to the third-stage prior as half-Gaussian.  This is almost correct, but does not account for the leading term of $\lambda^{-1}$, which is needed to cancel with the normalizing constant of the $\DE(0, \lambda^{-1})$ prior in the middle stage.
\end{example}

\subsection{The dual mixture representation}

Both examples suggest a natural statistical interpretation for the partial minimization step in $\lambda$: the profiling out of a nuisance parameter $\lambda$ in a hierarchical model where $(x \mid \lambda) \sim \DE(0, \lambda^{-1})$ and $p(\lambda) \propto \lambda^{-1} e^{\phi^{\star}(\lambda)}$.

We have referred to this interpretation as ``quasi-Bayesian'': although (\ref{eqn:jointgammalasso}) describes a typical hierarchical model, profiling out $\lambda$ makes no sense from a Bayesian perspective.  Instead, the natural approach starting from (\ref{eqn:jointgammalasso}) is to marginalize over $\lambda$.  Doing so would lead to a different estimator: if $p(x, \lambda)$ is a joint distribution having marginal $p(x) = \int p(x, \lambda) d \lambda$, the MAP estimate for $p(x)$ does not equal the $x$ ordinate of the joint MAP estimate for $p(x, \lambda)$ \citep[e.g.][]{ohagan:1976}.  For example, marginalizing over $\lambda$ in (\ref{eqn:jointgammalasso}) leads to the normal-exponential-gamma model \citep{griffin:brown:2005}, whereas profiling leads to the double-Pareto model.  These priors differ in functional form.

This raises several interesting questions.  Suppose that we begin with the model
$$
y \sim p(y \mid x) \; , \quad (x \mid \lambda) \sim \DE(x \mid 0, \lambda^{-1}) \; , \quad \lambda \sim p_V(\lambda) \, ,
$$
and estimate $x$ by profiling out $\lambda$ under the working prior $p_V(\lambda)$.  When does this operation correspond to marginalizing out $\lambda$ under some \textit{other} true prior $p_I(\lambda)$?  (The $I$ and $V$ stand for the integral and variational representation, respectively.) That is, given a specific penalty $\phi(|x|)$, for what priors $p_I(\lambda)$ and $p_V(\lambda)$ does the relation
\begin{equation}
\label{HD:exponential}
e^{-\phi(|x|)}  \propto \int_{\R^+}  \DE(x \mid 0, \lambda^{-1}) \ p_I(\lambda) d \lambda \propto \sup_{\lambda \geq 0} \left\{ \DE(x \mid 0, \lambda^{-1}) \ p_V(\lambda) \right \}
\end{equation}
hold? When it does, profiling with respect to $p_V(\lambda) \equiv \lambda^{-1} e^{\phi^{\star}(\lambda)}$ and marginalizing with respect to $p_I(x)$ may be thought of as dual operations to one another, and the priors themselves labeled as hierarchical duals.

The following result provides a partial answer to this question.  It shows that all penalty functions $\phi(x)$ with completely monotone derivatives yield true Bayesian posterior modes under a mixture-of-exponentials prior and have corresponding dual priors as in (\ref{HD:exponential}).  Therefore, although profiling would not initially seem to be a natural Bayesian operation, there are situations where profiling with respect to a working prior is equivalent to marginalizing under some true prior.

\begin{theorem}[Duality for exponential mixtures] 
\label{thm:dualexponential}
Suppose that a density $f(x)$ is a mixture of exponentials with mixing density $p_I(\lambda)$:
$$
f(x) = \int_0^{\infty} \lambda e^{-\lambda x} p(\lambda) d(\lambda) \, .
$$
Then $f(x)$ is also an envelope of exponentials as in (\ref{HD:exponential}), under the working prior $p_V(\lambda) \propto \lambda^{-1} e^{\phi^{\star}(\lambda)}$, where $\phi^{\star}(\lambda)$ is the concave conjugate of $\phi(x) = -\log f(x)$.   Conversely, suppose that $\phi(0) = 0$ and that $\phi(x)$ is nonnegative with completely monotone derivative.  Then
$$
e^{-\phi(x)} = \int_0^{\infty} e^{-\lambda x} d F(\lambda) \, 
$$
for some probability measure $F(\lambda)$.  If $F$ has a density $f(\lambda)$, the mixing measure in (\ref{HD:exponential}) is $p_I(\lambda) \propto \lambda^{-1} f(\lambda)$.
\end{theorem}

One caveat is that the implied prior $p_I(\lambda)$ in Theorem \ref{thm:dualexponential} need not be proper.  Nonetheless, there is a long tradition in Bayesian statistics of using improper priors \citep[see][for a review]{bergerBA2004}.  The resulting estimator is a valid posterior mode as long as the implied posterior density is bounded and proper.

\subsection{Equivalence of mixture and envelope representations}

The second (converse) statement of Theorem \ref{thm:dualexponential} is weaker than the forward direction in two senses.  First, the existence of a concave envelope representation (\ref{eqn:concavevariational}) is necessary but not sufficient for an integral representation to exist; this requires the additional condition that $\phi'(x)$ is completely monotone.  Second, even when the integral representation does exist, $F(\lambda)$ cannot be easily identified, except in special cases where the inverse Laplace transform of $e^{-\phi(x)}$ is available in closed form.

Luckily, we need not identify $p_I(\lambda)$ explicitly in order to see the operational equivalence of profiling (under $p_V$) and marginalizing (under $p_I$).  For example, consider the posterior distribution corresponding to the simple Gaussian model,
$$
q(x \mid y) \propto \exp \left\{ -\frac{1}{2} (y-x)^2 \right\} \int_{0}^{\infty} \lambda e^{-\lambda |x|} p_I(\lambda) d \lambda \, .
$$
This naturally suggests the following EM algorithm for computing the MAP estimate $\hat{x}$.
\begin{description}
\item[E step.] Compute the expected value of the complete-data log posterior from (\ref{HD:exponential}):
$$
Q(x \mid x^{(t)}) = E_{(\lambda \mid x)} \left\{   -\frac{1}{2} (y-x)^2  - \lambda |x| \right\} =  -\frac{1}{2} (y-x)^2  - \lambda^{(t)}  |x| \, ,
$$
where $\lambda^{(t)} = E(\lambda \mid x^{(t)})$.  This may be calculated using the relation
\begin{eqnarray*}
p'(x) &\propto& - \phi'(x) e^{-\phi(x)} = \int_{0}^{\infty} \frac{d}{dx} \lambda e^{-\lambda x} p_I(\lambda) d \lambda \\
&=& - \int_{0}^{\infty} \lambda \left[ \lambda e^{-\lambda x} p_I(\lambda) \right] d \lambda \, ,
\end{eqnarray*}
implying that
$$
\phi'(x) = \frac{\int_{0}^{\infty} \lambda \left[ \lambda e^{-\lambda x} p_I(\lambda) \right] d \lambda}
{\int_{0}^{\infty} \lambda e^{-\lambda x} p(\lambda) d \lambda} = E(\lambda \mid x) \, .
$$
This matches the profiling (or majorization) step in (\ref{eqn:soft.threshold}).

\item[M step.] Maximize $Q(x \mid x^{(t)})$:
$$
x^{(t+1)} = \arg \min_x \left\{  \frac{1}{2} (y-x)^2  + \lambda^{(t)} |x| \right\} =  S(y; \lambda^{(t)}) \, .
$$
This corresponds to the minimization step of the MM algorithm in (\ref{eqn:soft.threshold}).
\end{description}

In summary, the forward direction of Theorem \ref{thm:dualexponential} shows that any prior representable as a mixture of exponentials has a corresponding variational representation in terms of a dual function $\phi^{\star}(\lambda)$, and a working prior $p_V(\lambda) \propto \lambda^{-1} e^{\phi^{\star}(\lambda)}$.  This dual may be explicitly computed under far more general circumstances than those in which the mixing measure $p_I(\lambda)$ is known.

Moreover, if the penalty function satisfies the stated conditions of the theorem, then $f(x) = e^{-\phi(x)}$ is a mixture of exponentials, and profiling corresponds to the more stereotypically Bayesian operation of marginalizing, albeit under some other prior for $\lambda$.  This result provides a statistical interpretation of the majorization step---profiling a nuisance parameter in a hierarchical model---together with the notion of a dual prior as the $p_I(\lambda)$ that would yield the same estimator if one marginalized instead.

One useful point of comparison with our results is the monograph by \citet{wainwright:jordan:2008}.  They show that cumulant-generating functions in exponential families have the dual representation
$$
e^{A_y(\theta)} =  \int_{\mathcal{X}}  \exp \left\{ \theta^T \phi(x,y) \right\} \nu(dx) = \sup_{\mu} \exp \left\{ \theta^T \mu - A_y^{\star}(\mu) \right\} \, .
$$
This representation requires that the integrand in the mixture be a joint exponential family with sufficient statistic $\phi(x,y)$, where $x$ is the parameter being integrated out.  While neither our result nor theirs nests the other, they do share the same motivation of exploiting duality theory to show a connection between mixture and variational representations of statistical objective functions.

\section{Hierarchical duality: the Gaussian case}

\subsection{Random scale}

Similar duality results are also available for many conditionally Gaussian models.  We will begin with scale mixtures and then proceed to location mixtures and variance--mean mixtures. We will see that these all arise from applying the Fenchel--Moreau theorem to transformations of the original objective function.  This is analogous to the way in which Gaussian scale mixtures arise from applying the Bernstein--Widder theorem to transformations of a probability density function.

Consider any $\phi(x): \mathcal{R}^+ \to \mathcal{R}^+$ for which the function $\theta(x) = \phi(\sqrt{2x})$ is concave.  This class of penalties was studied by \citet{geman:reynolds:1992}, who used them to detect edges in blurred images.  By Lemma \ref{fact:fenchel}, we may write any such $\phi(x)$ as
\begin{equation}
\label{eqn:gemanreynolds}
\phi(x) = \theta(x^2/2) = \inf_{\lambda \geq 0} \left\{ \frac{\lambda}{2} x^2 - \theta^\star(\lambda) \right\} \, ,
\end{equation}
where $\theta^\star(\lambda)$ is the concave dual of $\theta(x)$.  Thus the prior is an envelope of normals with a random scale,
$$
e^{-\phi(x)} = \sup_{\lambda \geq 0} \left\{ \N(x \mid 0, \lambda^{-1}) \ p_V(\lambda) \right \} \, ,
$$
where $p_V(\lambda)$ may be expressed in terms of the concave dual for $\theta(x)$.  Combining this with some basic facts about normal scale mixtures leads to the following duality result.

\begin{theorem}
\label{thm:scalemixtureduals}
Suppose that $p(x) \propto e^{-\phi(x)}$ is symmetric in $x$, and let $\theta(x) = \phi(\sqrt{2x})$ for positive $x$.  Suppose that $\theta'(x)$ is completely monotone.  Then $p(x)$ has both a mixture and envelope representation in terms of a conditionally Gaussian model with a random scale:
\begin{equation}
\label{HD:normalscale}
f(x) = e^{-\phi(x)}  \propto \int_{\R^+}  \N(x \mid 0, \lambda^{-1}) \ p_I(\lambda) d \lambda \propto \sup_{\lambda \geq 0} \left\{ \N(x \mid 0, \lambda^{-1}) \ p_V(\lambda) \right \} \, ,
\end{equation}
where the working variational prior is $p_V(\lambda) \propto \lambda^{-1/2} e^{\theta^\star(\lambda)}$.  Moreover, any optimal value of $\lambda$ in the envelope representation satisfies $\hat \lambda(x) \in \partial \theta(x^2/2)$, or
$$
\hat \lambda(x) = \frac{\phi'(x)}{x} 
$$
whenever $\phi$ is differentiable.
\end{theorem}
Complete monotonicity of $\theta'(x)$ is sufficient to ensure that $\theta(x)$ is concave, although the converse does not hold.  Thus the class of priors representable as scale mixtures of normals is a strict subset of those representable as envelopes of normals.  This fact was also observed by \citet{palmer:etal:2006}.  

As in Section \ref{section:exponential}, both $p_I(\lambda)$ and $p_V(\lambda)$ have Bayesian interpretations as priors for $\lambda$ in a hierarchical model, though in this case $p(x \mid \lambda)$ is Gaussian with variance $\lambda^{-1}$.  Again as before, both interpretations lead to the same iterative algorithm.  Consider the envelope representation first.  Following (\ref{eqn:gemanreynolds}), consider the problem
$$
\hat{x} = \arg \min_{x} \left\{ \frac{1}{2}(y-x)^2 + \phi(x) \right\} = \arg_x \min_{x, \lambda} \left\{ \frac{1}{2}(y-x)^2 + \frac{\lambda}{2} x^2 - \theta^\star(\lambda) \right\} \, .
$$
The second equality leads to the following iterative scheme:
\begin{equation}
\label{eqn:scalemixtureEM}
\begin{aligned}
x^{(t+1)} &=  \arg \min_x \left\{\frac{1}{2}(y-x)^2 + \frac{1}{2} \lambda^{(t)} x^2 \right\} \\
\lambda^{(t+1)} &= \arg \min_\lambda \left\{  \frac{1}{2} \lambda [x^{(t+1)}]^2 - \theta^\star(\lambda) \right\} = \frac{\phi' \big( x^{(t+1)} \big) }{x^{(t+1)}}   \, .
\end{aligned}
\end{equation}
If $\phi$ is not differentiable, we instead use any element of the superdifferential $\partial \theta(x^2/2)$ in the second step, where $\theta(x) = \phi(\sqrt{2x})$.

Next, consider the mixture representation and the corresponding EM algorithm.  Let $\hat{\lambda}(x) = \E(\lambda \mid x)$ be the expected value of $\lambda$, given $x$, under the prior $p_I(\lambda)$.  The complete-data log posterior arising from (\ref{HD:normalscale}) is
$$
Q(x \mid x^{(t)}) = \E_{(\lambda \mid x^{(t)})} \left\{ -\frac{1}{2}(y-x)^2 - \frac{\lambda}{2} x^2 \right\}   =  -\frac{1}{2}(y-x)^2 - \frac{\hat{\lambda}(x^{(t)})}{2} x^2\, ,
$$
ignoring constants not depending on $x$.  We can compute $\E(\lambda \mid x)$ using the identity
\begin{eqnarray}
\frac{\dd p(x)}{\dd x} &\propto& -\phi'(x) e^{-\phi(x)} = \int_{\R^+} \frac{\dd}{\dd x} \N(x \mid 0, \lambda^{-1} ) p(\lambda) \dd \lambda \label{eqn:masreliezstyle} \\
&=& -x \int_{\R^+} \lambda \ \N(x \mid 0, \lambda^{-1}) p(\lambda) \dd \lambda \nonumber \, .
\end{eqnarray}
Dividing through by $e^{-\phi(x)} = \int_{\R^+} \N(x \mid 0, \lambda^{-1} ) p(\lambda) \dd \lambda$ gives an expression for the conditional moment needed in an EM algorithm,
$$
E(\lambda \mid x) = \frac{\phi'(x)}{x} \, .
$$
which leads to the same update rule as (\ref{eqn:scalemixtureEM}).  Note that this argument explicitly requires the differentiability of $\phi$, rather than merely the existence of a subdifferential.  This parallels the stronger conditions (complete monotonicity of $\theta(x)$) needed for the mixture representation to exist in the first place.

\begin{example}[Binomial logit]
\normalfont
\label{ex:logitscale}
Consider a simple binomial model parameterized by the log odds of success, as in a logistic regression model: $(y \mid x) \sim \mbox{Binom}\{m, w(x)\}$, where $m$ is a fixed number of trials, and $w(x) = 1/(1+e^{-x})$.  The negative log likelihood is
$$
\begin{aligned}
l(x) &= m \log \{ 1 + \exp(x)\} - y x  \\
&= m \log \{ \exp(x/2) + \exp(-x/2) \} - (y-m/2)x \\
&= m \log \cosh(x/2) - \kappa x - \log 2 \, ,
\end{aligned}
$$
where $\kappa = y-m/2$.
The function $\log \cosh(x/2)$ satifies the conditions of Theorem \ref{thm:scalemixtureduals}:
\begin{equation}
\label{eqn:logitlx}
\log \cosh(x/2) = \inf_{\lambda \geq 0} \left\{ \frac{\lambda}{2}x^2 - \theta^\star(\lambda) \right\} \, .
\end{equation}
Therefore we may write the logit likelihood as both a mixture and an envelope of a conditionally Gaussian model with a random scale:
\begin{equation}
\frac{ \{e^x\}^y } {\{ 1+e^x \}^m}  \propto e^{\kappa x} \int_{\R^+}  \N(x \mid 0, \lambda^{-1}) \ p_I(\lambda) d \lambda = e^{\kappa x} \sup_{\lambda \geq 0} \left\{ \N(x \mid 0, \lambda^{-1}) \ p_V(\lambda) \right \} \, .
\end{equation}

Each representation has been used independently of the other, without their connection being appreciated.  \cite{jaakkola:jordan:2000} used the envelope representation to construct a variational-Bayes estimate in the logistic-regression model.  Meanwhile, \citet{polson:scott:windle:2012a} used the mixture representation to construct a Gibbs sampler, and identified the mixing distribution $p_I(\lambda)$ as an infinite convolution of gammas known as the Polya-Gamma distribution.  The minimizing value of $\lambda$ in (\ref{eqn:logitlx}) is the same as the conditional moment $\E(\lambda \mid x)$ in the Polya-Gamma mixture representation:
$$
\hat \lambda(x) = E_{PG}(\lambda \mid x) = \frac{m}{2x} \tanh(x/2) \, .
$$

\end{example}

\subsection{Random location}

The following result from \citet{geman:yang:1995} establishes a class of priors that may be written as envelopes of normals with a random location parameter.  They refer to this as the family of half-quadratic regularizers. 
\begin{lemma}[Geman and Yang, 1995]
\label{lemma:gemanyang}
\begin{enumerate}[(A)]
\item Let $\phi(x)$ be a function such that $\theta(x) = \frac{1}{2} x^2 - \phi(x)$ is closed and convex.  Equivalently, let $\psi(\lambda)$ be a function such that $\eta(\lambda) = \frac{1}{2}\lambda^2 + \psi(\lambda)$ is closed and convex.  If either condition holds, then the following dual relationship holds:
\begin{eqnarray*}
\phi(x) &=&  \inf_{\lambda \in \mathcal{R}} \left\{ \frac{1}{2} (x - \lambda)^2 + \psi(\lambda)  \right\} \\
\psi(\lambda) &=&  \sup_{x \in \mathcal{R}} \left\{ - \frac{1}{2} (x - \lambda)^2 + \phi(x)  \right\}  \, .
\end{eqnarray*}
\item Any minimizing value of $\lambda$ in the expression for $\phi$ satisfies
$$
\hat{\lambda}(x) \in \partial \left\{\frac{1}{2} x^2 - \phi(x)\right\} \, .
$$ In the case of a differentiable $\phi$, this becomes $\hat{\lambda}(x) = x - \phi'(x)$.
\end{enumerate}
\end{lemma}

We give a proof in the appendix.  But the idea is simply to apply Lemma \ref{fact:fenchel} to $\theta(x)$ or $\eta(\lambda)$, thereby establishing that these functions form a Legendre pair.  We now state our duality result for Gaussian models with random locations. 

\begin{theorem}
\label{theorem:locationmixture}
Suppose that $f(x) = e^{-\phi(x)}$, $\phi(x) > 0$, satisfies the following conditions:
\begin{enumerate}
\item $\int_\R f(x) dx < \infty$.
\item $\phi(x)$ has continuous derivatives of all order.
\item The series $\sum_{k=0}^{\infty} \frac{(-1)^k}{4^k k!} f^{(2k)}(x)$ converges uniformly to a non-negative value.
\item The function $\theta(x) = \frac{1}{2} x^2 - \phi(x)$ is closed and convex, with dual function $\theta^{\star}(\lambda)$.
\end{enumerate}
Then $f(x)$ has both a mixture and an envelope representation as a normal location model:
$$
p(x) \propto e^{-\phi(x)} = \int_{-\infty}^{\infty} \N(x \mid \lambda, 1) \ p_I(\lambda) d \lambda =  \sup_{\lambda \in \mathcal{R}} \left\{ \N(x \mid \lambda, 1) \ p_V(\lambda) \right\} \, ,
$$
where the variational prior for $\lambda$ is
\begin{eqnarray*}
p_V(\lambda) &=& e^{-\psi(\lambda)} \; , \quad  \psi(\lambda) = \theta^{\star}(\lambda) - \frac{1}{2} \lambda^2 \, .
\end{eqnarray*}
\end{theorem}

This is a weaker duality result than both of the previous two: the conditions under which both a mixture and an envelope representation exist are more restrictive than for the case of random-scale models.  Moreover, neither class is strictly broader than the other.  Condition (4) must hold in order for an envelope representation to exist, and this will fail for many Gaussian location mixtures, such as those with many  modes.  As for the other direction, the following result from \citet{dasgupta:1994} shows that many familiar distributions with easily derived envelope representations cannot be represented as Gaussian location mixtures.

\begin{theorem}[Dasgupta, 1994]
Suppose that $p(x)$ is a Gaussian scale mixture:
$$
p(x) = \int \N(x \mid 0, \sigma^2) \ p(\sigma^2) \ d \sigma^2 \, .
$$
Then $p(x)$ may also be represented as a Gaussian location mixture if and only if the mixing distribution $p(\sigma^2)$ places no mass on the interval $[0,1)$.
\end{theorem}
This rules out a large number of common families with simple Gaussian envelope representations, including the Student $t$, double exponential, and logistic distributions.   To see this, observe that all three are known to be Gaussian scale mixtures, and the corresponding mixing distributions for the variance all place positive mass on $[0,1)$. They therefore cannot be Gaussian convolutions.

It is evident from these results that the envelope representation will be more useful for most practical problems.  We now give three examples.

\begin{example}[Limited-translation rule]
\normalfont
Consider the limited-translation rule of \citet{efron:morris:1972}, which corresponds to the loss function $\phi(x) = \min(1, x^2/2)$.  The implied prior is Gaussian near the origin, but improper.  Simple algebra yields the $\psi(\lambda)$ function in the envelope representation of Lemma \ref{lemma:gemanyang}:
$$
\psi(\lambda) = 
\begin{cases}
1 - \frac{(\sqrt{2} - \lambda)^2}{2} &\mbox{if } 0 \leq \lambda \leq \sqrt{2} \\ 
1 & \mbox{if } \lambda > \sqrt{2} \, .
\end{cases} 
$$
\citet{efron:morris:1972} used the rule as a compromise between Stein's estimator and the maximum-likelihood estimate.  The idea is to limit the risk associated with individual components of a multivariate location parameter, while giving up only a small fraction of the reduction in total risk given by Stein's rule.
\end{example}

\begin{example}[A non-sparse penalty]
\normalfont
Lemma \ref{lemma:gemanyang} offers the option of specifying a prior directly via the dual function $\psi(\lambda)$ in the envelope representation.  Consider the function
$$
\psi(\lambda) = \frac{\lambda}{2(1+\lambda)} \
$$
on $\lambda \in \R^+$.  It is easy to check that $\frac{1}{2}\lambda^2 + \psi(\lambda)$ is a convex function on $\R^+$, and so we may define
$$
\phi(x) =  \inf_{\lambda \geq 0} \left\{ \frac{1}{2}(x-\lambda)^2 + \frac{\lambda}{2(1+\lambda)} \right\} \, .
$$
Although $\phi(x)$ lacks a simple closed form, it may be evaluated numerically.  It is neither globally convex nor concave, behaving like a quadratic function near the origin and like the function $\psi$ itself for large arguments  \citep{geman:yang:1995}.  It is interesting primarily because it will behave like ridge regression near the origin, and thus will not induce a sparse estimator.  The desirability of this property will depend on context.  For example, \citet{leeb:potscher:2005} discuss some of the potential problems with sparse estimators, relating them to the same unbounded risk property that arises with the use of Hodges' thresholding estimator.
\end{example}

\begin{example}[Binomial logit, part 2]
\normalfont
Again suppose that $(y \mid x) \sim \mbox{Binom}\{m, w(x)\}$, with $m$ fixed and $w(x) = 1/(1+e^{-x})$.  Let $\kappa = y-m/2$.  Write the negative log likelihood as
$$
\begin{aligned}
l(x) &= m \log \cosh(x/2) - \kappa x - \log 2 \, .
\end{aligned}
$$
The function $\log \cosh(x/2)$ satisfies the conditions of Lemma \ref{lemma:gemanyang}, and so we may write
$$
l(x) = m \inf_{\lambda} \left\{\frac{1}{2}(x-\lambda)^2  + \psi(\lambda)  \right\} - \kappa x \, ,
$$
up to a constant.  For fixed $\lambda$, this is quadratic in $x$. This gives an alternate conditionally Gaussian representation of the logistic-regression model to that of Example \ref{ex:logitscale}.


\end{example}

In each case, we recognize the representation for $f(x)$ as the Moreau envelope of some function $\psi(\lambda)$, and the partial minimization step in $x$ as the proximal mapping of that function.  Though this equivalence does not help us actually compute the updates, it does connect the idea of a Gaussian envelope representation with familiar ideas from variational analysis.

\subsection{Variance-mean envelopes}

We now generalize the results of \citet{geman:reynolds:1992} and \citet{geman:yang:1995} to the case of variance-mean envelopes of Gaussians.  These are analogous to the widely studied class of variance-mean mixtures \citep[e.g.][]{barndorffnielsen:1978,bn:kent:sorensen:1982,Polson:Scott:2011a}.  Because they are usually asymmetric, such representations are typically useful for handling likelihood and pseudo-likelihood functions rather than penalties.

\begin{theorem}
\label{thm:varmeanenvelope}
Let $f(x)$ be some function, and suppose there exists a $\kappa \in \R$ for which $g(x) = f(x) + \kappa x$ is symmetric in $x$ and has the property that $\theta(x) = g(\sqrt{2x})$ is concave on $\R^+$.  Let $\theta^{\star}(\lambda)$ be the concave dual for $\theta(x)$.  Then $p(x) \propto e^{-f(x)}$ has an envelope representation as a variance--mean normal distribution with drift parameter $\kappa$:
$$
p(x) \propto e^{-f(x)} =  \sup_{\lambda} \left\{   \N(x \mid \kappa \lambda^{-1}, \lambda^{-1}) \ \lambda^{-1/2} e^{\psi(\lambda)} \right\} \, ,
$$
where $\psi(\lambda) = \theta^{\star}(\lambda) + \kappa^2/2\lambda$.  Moreover, any optimal value of $\lambda$, as a function of $x$, satisfies $\hat \lambda(x) \in \partial \theta(x^2/2)$.  In the case where $f$ is differentiable, this becomes
$$
\hat \lambda(x) = \frac{f'(x) + \kappa}{x} \, .
$$
\end{theorem}

\begin{example}[Quantile regression]
\label{example:quantreg}
\normalfont
Choose $q \in (0,1)$ and let  $l(x) = |x| + (2q-1)x$.  This is the hinge loss function, and is used in quantile regression for the $q$th quantile \citep{koenker:2005}.  \citet{li:xi:lin:2010} represent this as a pseudo-likelihood involving the asymmetric Laplace distribution.  We derive the corresponding envelope representation as a variance-mean Gaussian.

Let $\kappa = 1- 2q$.  Then $g(x) = l(x) + \kappa x$ is clearly symmetric in $x$, and is concave in $x^2$, and the conditions of the theorem apply.  We thus have
$$
l(x) = \inf_{\lambda \geq 0} \left\{ \frac{\lambda}{2}\left( x - \frac{1-2q}{\lambda} \right)^2 - \psi(\lambda) \right\} \, .
$$
In this case $\theta(x) = \sqrt{2x}$, which has concave dual $\theta^\star(\lambda) = -1/(2\lambda^2)$.  Thus
$$
\psi(\lambda) = \frac{\kappa^2}{2\lambda} - \frac{1}{2 \lambda^2} \, ,
$$
and the conditional mode for $\lambda$ is
$$
\hat{\lambda}(x) = \mbox{sgn}(x)/x \, .
$$

\end{example}

\section{Multivariate envelopes}

\subsection{A generalization of the random-location case}

So far we have appealed to mixture and envelope representations of univariate distributions, which must be applied component-by-component in multivariate problems.  But the following result provides a multivariate generalization of \citet{geman:yang:1995} for a wide class of priors and likelihoods that may be represented in terms of a conditionally Gaussian location model.

\begin{theorem}
\label{lemma:lipschitzenvelope}
Let $p(x)$ be a likelihood or prior distribution for $x \in \R^d$, and let $f(x) = -\log p(x)$. Suppose that $\nabla f(x)$ exists and satisfies
$$
\enorm{ \nabla f(x) - \nabla f(y) } \leq \frac{1}{c} \enorm{x-y} 
$$
for some $c$, and for all $x,y$.  Then $p(x)$ has an envelope representation in terms of a multivariate normal location model:
$$
p(x) = \sup_{\lambda \in \R^d} \left\{ \N(x \mid c \lambda, c I) \ p_V(\lambda) \right\} \, .
$$
If $f(x)$ is differentiable, then the optimal vector $\lambda$, as a function of $x$, is
$$
\hat \lambda(x) = c^{-1}x - \nabla f(x) \, .
$$
Here $p_V(\lambda) \propto \exp\{\psi(\lambda)\}$ where $\psi$ is the conjugate function of $f(x)$.

\end{theorem}
The key requirement is that the function being represented must have a Lipschitz-continuous gradient with Lipschitz constant $L=1/c$.  For example, in a least-squares problem where $l(x) = \frac{1}{2} \enorm{y - Ax}^2$, the minimal Lipschitz constant for $l(x)$ is $l_d$, the maximum eigenvalue of $A^T A$.  In logistic regression, the minimal L is $l_d/4$.  One example of a likelihood where this condition fails to hold is in Poisson regression with the canonical log link function.

To see the utility of this representation, consider a Bayesian model of the form (\ref{eqn:bayesversion}), where $p(y \mid x) \propto e^{-l(x)}$ satisfies the conditions of Theorem \ref{lemma:lipschitzenvelope} for some constant $c$.  Rewrite $l(x)$ as
$$
l(x) = \inf_{\lambda \in \R^d} \left\{ \frac{1}{2c} \enorm{x-c\lambda}^2 + \psi(\lambda) \right\}
$$
with $\psi(\lambda)$ given by the lemma.  To find the posterior mode, we therefore need to solve the problem
$$
\begin{aligned}
& \underset{x, \lambda}{\text{minimize}}
& & \frac{1}{2c} \enorm{x - c \lambda}^2 + \psi(\lambda) + \phi(x) \, ,
\end{aligned}
$$
which is clearly equivalent to solving problem (\ref{eqn:canonicalform}).

This leads to the following iterative algorithm, which uses only gradient evaluations of the likelihood.
\begin{equation}
\label{eqn:proxgradlipschitz}
\begin{aligned}
\lambda^{(t+1)} &= a^{-1} x^{(t)} - \nabla l(x^{(t)}) \\
x^{(t+1)} &= \arg \min_x \left\{ \frac{1}{2a} \enorm{x - a \lambda^{(t+1)}}^2 + \phi(x) \right\} \, .
\end{aligned}
\end{equation}
We recognize the second step as the proximal operator of the penalty $\phi(x)$, evaluated at $a \lambda^{(t+1)}$.  For many penalties this operator is computationally negligible, and has a closed-form solution.  Alternatively, it can be solved by appyling the method of Section \ref{section:exponential} to each component of $x$, assuming that $\phi(x)$ is separable.

\subsection{Example: binomial logit, part 3}

As an example, return again to the binomial logit model, this time with design matrix $A$ and $d$-dimensional regression vector $x$.  Suppose that $x$ is given a double-Pareto prior.  Here the MAP estimate is the solution to the problem
$$
\begin{aligned}
& \underset{x \in \R}{\text{minimize}}
& &l(x) + \gamma \phi(x) = \sum_{i=1}^n \left\{ m_i \log(1+e^{a_i^Tx}) - y_i a_i^T x \right\} + \gamma \sum_{j=1}^d \log(1 + |x_j|/a) \, .
\end{aligned}
$$
As before, there are $m_i$ trials and $y_i$ successes at each design point $a_i$. The log likelihood satisfies the conditions of Theorem \ref{lemma:lipschitzenvelope} with $a = 4/l_d$, where $l_d$ is the maximum eigenvalue of $A^T A$.  Therefore the $\lambda$ update in (\ref{eqn:proxgradlipschitz}) becomes
$$
\lambda^{(t+1)} = \frac{l_d}{4} x^{(t)} - A^T r(x^{(t)}) \; , \quad [r(x)]_i = m_i \left(\frac{1}{1+e^{-a_i^Tx}} \right) - y_i \, .
$$

Meanwhile, we can evaluate the $x$ update in closed form.  The overall problem clearly separates component by component, and so we must solve the scalar problem
$$
\begin{aligned}
& \underset{x \in \R}{\text{minimize}}
& & \left\{  \frac{s}{2} (x-u)^2 + \gamma \log(1 + |x|/a)  \right\}
\end{aligned}
$$
This is clearly equivalent to the problem
\begin{equation*}
\begin{aligned}
& \underset{x \in \R,z \geq 0}{\text{minimize}}
& & \left\{  \frac{s}{2} (x-u)^2 + \gamma \log(1 + z/a)  \right\} \\
& \text{subject to}
& & z = |x| \, .
\end{aligned}
\end{equation*}

First consider the case $u \geq 0$.  In this case, the best choice of $x$ is clearly nonnegative and we may optimize over $x \geq 0$.  Likewise, if $u < 0$, the best choice of x is nonpositive, and we may optimize over $x \leq 0$.  In either case, with some algebra we reach an equivalent problem that may be written as
\begin{equation*}
\begin{aligned}
& \underset{z \in \R,  y \geq 0}{\text{minimize}}
& & \left\{ \frac{s}{2} (x-|u|)^2  + \gamma \log(1 + z/a) \right\}
\\
& \text{subject to}
& & \sgn(x) = \sgn(u) \; , |x| = z \,,
\end{aligned}
\end{equation*}
where the optimal value of $z$ provides the solution to the original problem.  This is differentiable in $z$ and therefore easily solved for both $z$ and $x$:
$$
\prox_{\phi/s}(u) = \frac{\sgn(u)}{2} \left\{ |u| - a + \sqrt{ (a- |u|)^2 + 4 d(u) } \right\} \; ,  \quad d(u) = (a|u| - \gamma/s)_+ \, .
$$

\subsection{The connection with the proximal gradient method}

We now show that the multivariate Gaussian envelope in Theorem \ref{lemma:lipschitzenvelope} provides a statistically meaningful missing-data interpretation for the proximal gradient algorithm, a widely used tool in signal processing \citep[see, e.g.][]{combettes:pesquet:2011}.  The proximal gradient method is usually motivated as an algorithm for finding the fixed point of a forward-backward operator derived from standard optimality conditions in subdifferential calculus.

We first sketch the operator-theoretic justification of the proximal-gradient algorithm, before making the connection with Gaussian location envelopes.  Suppose that $l(x)$ is differentiable but that $\phi(x)$ is not, and let $\partial$ be the subdifferential operator.  A necessary and sufficient condition for $x^{\star}$ to be the solution to (\ref{eqn:canonicalform}) is that
\begin{equation}
\label{eqn:subdiffproxgrad}
0 \in \partial \left\{ l(x) + \phi(x)\right\} = \nabla l(x) + \partial \phi(x) \, ,
\end{equation}
the sum of a point $\nabla l(x)$ and a set $\partial \phi(x)$ which is nonempty under quite general conditions.  We will use this fact to characterize $x^{\star}$ as the fixed point of the following operator:
$$
x^{\star} = \prox_{\gamma \phi}\{ x^{\star} - \gamma \nabla f(x^{\star}) \} \, .
$$
To see this, let $I$ be the identity operator.  Observe that finding the point $x^{\star}$ satisfying optimality condition (\ref{eqn:subdiffproxgrad}) is equivalent to finding the point $x^{\star}$ such that
$$
\begin{aligned}
0 &\in \gamma \nabla f(x^{\star}) - x^{\star} + x^{\star} + \gamma \partial \phi(x^{\star}) \\
x^{\star} - \gamma \nabla l(x^{\star}) &\in x^{\star} + \gamma \partial \phi(x^{\star}) \\
(I-\gamma \nabla f)x^{\star} &\in (I +  \gamma \partial \phi) x^{\star} \\
x^{\star} &= (I +  \gamma \partial \phi)^{-1} (I-\gamma \nabla f) x^{\star} \\
&=  \prox_{\gamma \phi} (I-\gamma \nabla f) x^{\star} \, .
\end{aligned}
$$
The penultimate line say that $x^{\star}$ is the fixed point of an operator defined by composing the two operators on the right-hand side.  The final line appeals to the fact that the proximal operator is the resolvent of the subdifferential operator: $\prox_{\gamma \phi}(x) = (I + \gamma \partial \phi)^{-1} (x)$.  Thus to find the solution, we repeatedly apply this operator to find $x^\star$ as a fixed point:
$$
x^{(t+1)} = \prox_{\gamma^k \phi}\{ x^{(t)} - \gamma^{(t)} \nabla f(x^{(t)}) \} \, ,
$$
for appropriate step size $\gamma^{(t)}$.

Now return to the two steps in (\ref{eqn:proxgradlipschitz}).  If we substitute the $\lambda$ update directly into the expression for the $x$ update, we have
$$
\begin{aligned}
x^{(t+1)} &=  \arg \min_x \left\{ \frac{1}{2a} \left\Vert  x - a  \left[ a^{-1} x^{(t)  } - \nabla l(x^{(t)}) \right]     \right\Vert_2^2 + \phi(x) \right\} \\
&=  \prox_{a \phi}\{  x^{(t)} - a \nabla l(x^{(t)}) \}\, ,
\end{aligned}
$$
which is precisely the proximal gradient method with step size $a$.  

When applied to least-squares problems, the proximal-gradient method is often called iterative shrinkage thresholding (IST).  \cite{fig:nowak:2003} provide an EM interpretation of this algorithm, but the interpretation does not carry through in the case of a non-Gaussian likelihood.  Our envelope representation is a different kind of missing-data argument, and applies to any log likelihood with a Lipschitz-continuous gradient.

\section{Statistical applications}
\label{sec:applications}

\subsection{Robust fused lasso}

Suppose we observe data $y_i = f(x_i) + e_i$ where $f(x)$ is piecewise constant.  The fused lasso \citep{tibs:fusedlasso:2005} involves estimating $f(x_i) \equiv \beta_i$ at the input points by solving the following optimization problem:
$$
\begin{aligned}
& \underset{\beta \in \R^d}{\text{minimize}}
& & \frac{1}{2} \Vert y - \beta \Vert_2^2 + \lambda \Vert D^{(1)} \beta \Vert_1 \, ,
\end{aligned}
$$
where $D^{(1)}$ is the matrix encoding the first differences in $\beta$:
\begin{equation}
\label{eqn:fusedlassoD}
D^{(1)} =\left(\begin{array}{rrrrrr}
1 & -1 & 0 & 0 & \mathbf{\cdots} & 0\\
0 & 1 & -1 & 0 & \cdots & 0\\
\vdots &  &  &  & \ddots & \vdots \\
0 & \cdots &  & 0 & 1 & -1
\end{array}\right).
\end{equation}

\begin{figure}
\begin{center}
\includegraphics[width=6in]{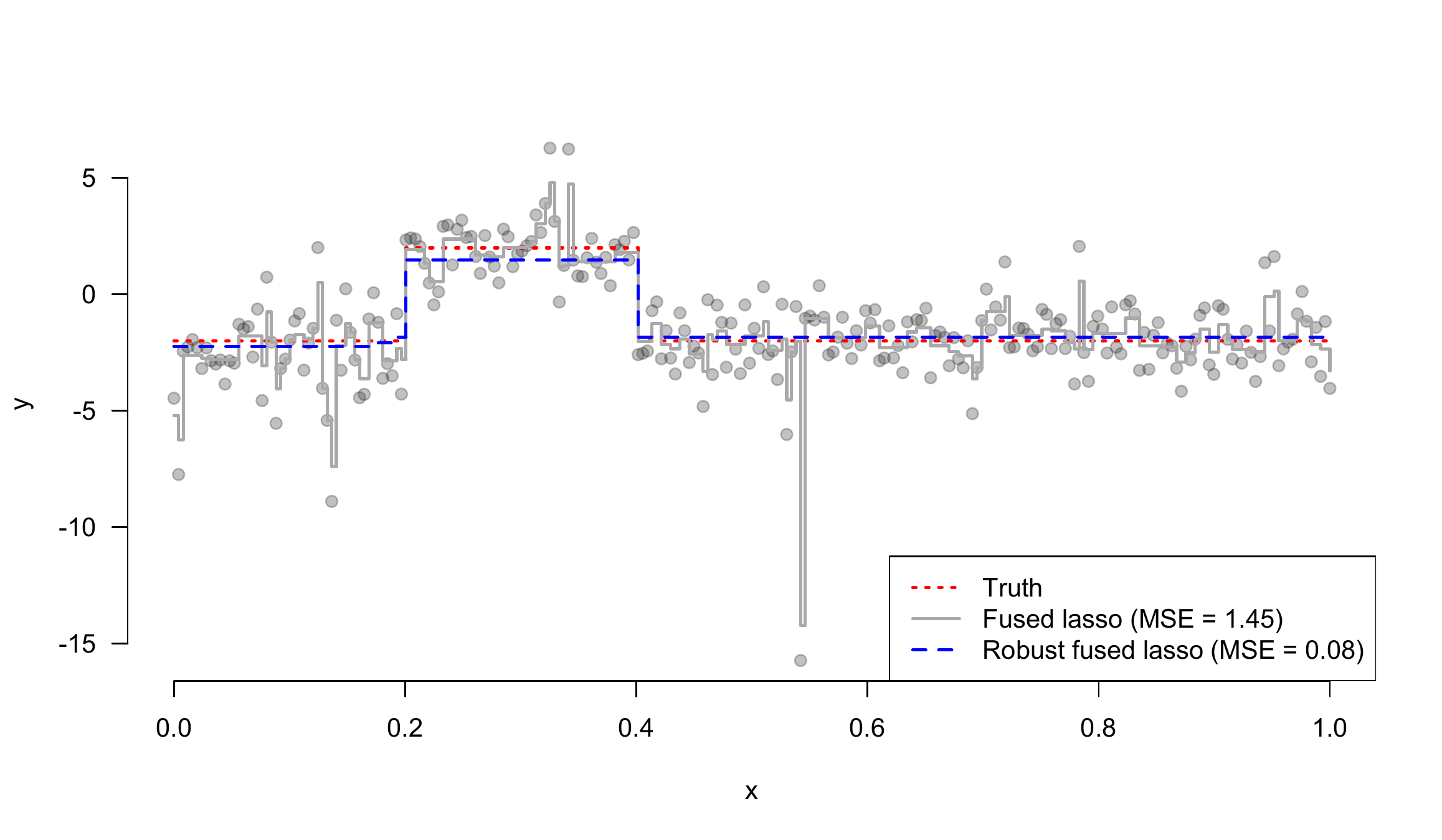}
\caption{\label{fig:rfl} Robust fused lasso fit with Huber loss, versus ordinary fused lasso.  For both procedures, the penalty parameter was chosen to minimize AIC, treating the loss function as a negative log likelihood.}
\end{center}
\end{figure}

With Lemma \ref{lemma:gemanyang} in mind, we implemented an ``outlier robust'' version of the fused lasso by using Huber loss rather than $\ell^2$ loss:
\begin{equation}
\label{eqn:robustFL}
\begin{aligned}
& \underset{\beta \in \R^d}{\text{minimize}}
& & \sum_{i=1}^n H(y_i - \beta_i) + \lambda \Vert D^{(1)} \beta \Vert_1 \, ,
\end{aligned}
\end{equation}
where
$$
H(x) = \left\{
\begin{array}{l l}
x^2/2 & \mbox{if $|x| < 1$} \, , \\
|x| - 1/2 & \mbox{if $|x| \geq 1$} \, .
\end{array}
\right.
$$
It is easily verified that $H(x)$ satisfies the conditions of Lemma \ref{lemma:gemanyang}, and that the quantity
$$
\hat{u}(x) =
\left\{
\begin{array}{l l}
0 & \mbox{if $|x| < 1$} \, , \\
x - \sgn(x) & \mbox{if $x \geq 1$} \, 
\end{array}
\right.
$$
lies in the subdifferential, evaluated at $x$, of the function $\theta(x) = x^2/2 - H(x)$.  This leads to a simple iterative scheme for solving (\ref{eqn:robustFL}):
$$
\begin{aligned}
r_i^{(t)} & = y_i - \beta_i^{(t-1)} \\
u_i^{(t)} & = \left\{
\begin{array}{l l}
0 & \mbox{if $|r_i^{(t)}| < 1$} \, , \\
r_i^{(t)} - \sgn(r_i^{(t)}) & \mbox{otherwise.} \, 
\end{array}
\right. \\
\beta^{(t)} & = \arg \min_\beta \frac{1}{2} \Vert(y-u^{(t)}) - \beta \Vert_2^2 +  \lambda \Vert D^{(1)} \beta \Vert_1 \, .
\end{aligned}
$$
The $\beta$ step is an ordinary fused lasso problem with working response vector $y-u^{(t)}$.  We solve this subproblem using the dynamic programming algorithm of \citet{johnson:2013}, as implemented in the \verb|glmgen| R package \citep{glmgen:2014}.

To illustrate the method, we simulated 250 observations from the true function shown in Figure \ref{fig:rfl}, in which the residuals were generated from a standard $t$ distribution with 3 degrees of freedom.  We ran both the ordinary fused laso and the robust fused lasso, in each case choosing lambda to minimize AIC across a grid of values.  For the Huber loss, we operationally defined AIC as twice the Huber loss at the optimum, plus twice the number of distinct levels of the fitted function.

As Figure \ref{fig:rfl} shows, the ordinary fused lasso, even with the AIC-optimal choice of $\lambda$, performs poorly in the presence of idiosyncratic large residuals.  The robust version has a much smaller reconstruction error.  This excellent performance comes at little computational cost: fitting the robust model across a grid of 100 lambda values took less than 0.2 seconds on an ordinary Apple laptop.

\subsection{Nonlinear quantile regression via trend filtering}

Polynomial trend filtering \citep{kim:boyd:etal:2009,tibs:2014a} is a recently proposed method for piece-wise polynomial curve-fitting, where the knots and the parameters are chosen adaptively.  Specifically, suppose we have observations $y_i$ observed on a regular grid $x_1, \ldots, x_n$, where $y_i = f(x_i) + e_i$ for some unknown function $f$.  The trend filtering estimator of order $k$ is the solution to the problem
$$
\begin{aligned}
& \underset{\beta \in \R^d}{\text{minimize}}
& & \frac{1}{2} \Vert y - \beta \Vert_2^2 + \lambda \Vert D^{(t+1)} \beta \Vert_1 \, ,
\end{aligned}
$$
where $D^{(t+1)}$ is the discrete difference operator of order $k+1$.  For $k = 0$ this matrix is the first-difference matrix from Equation (\ref{eqn:fusedlassoD}), and the model reduces to the fused lasso.  For $k \geq 1$ this matrix is defined recursively as $D^{(t+1)} = D^{(1)} D^{(t)}$, where $D^{(1)}$ is of the appropriate dimension.  Intuitively, the trend-filtering estimator is similar to an adaptive spline model: it penalizes the discrete derivative of order $k$, resulting in piecewise polynomials of higher degree for larger $k$.

The solution to the trend-filtering problem will fit a smooth function to the conditional mean of $y$ given $x$.  To illustrate our framework, we propose the trend-filtering quantile-regression estimator:
$$
\begin{aligned}
& \underset{\beta \in \R^d}{\text{minimize}}
& & \sum_{i=1}^n \{ |y_i - \beta_i| + (2q-1)(y_i - \beta_i) \} + \lambda \Vert D^{(t+1)} \beta \Vert_1 \, .
\end{aligned}
$$
The solution to this optimization problem will provide a nonparametric estimate of the $q$th quantile function for $y$ given $x$.

\begin{figure}
\begin{center}
\includegraphics[width=6in]{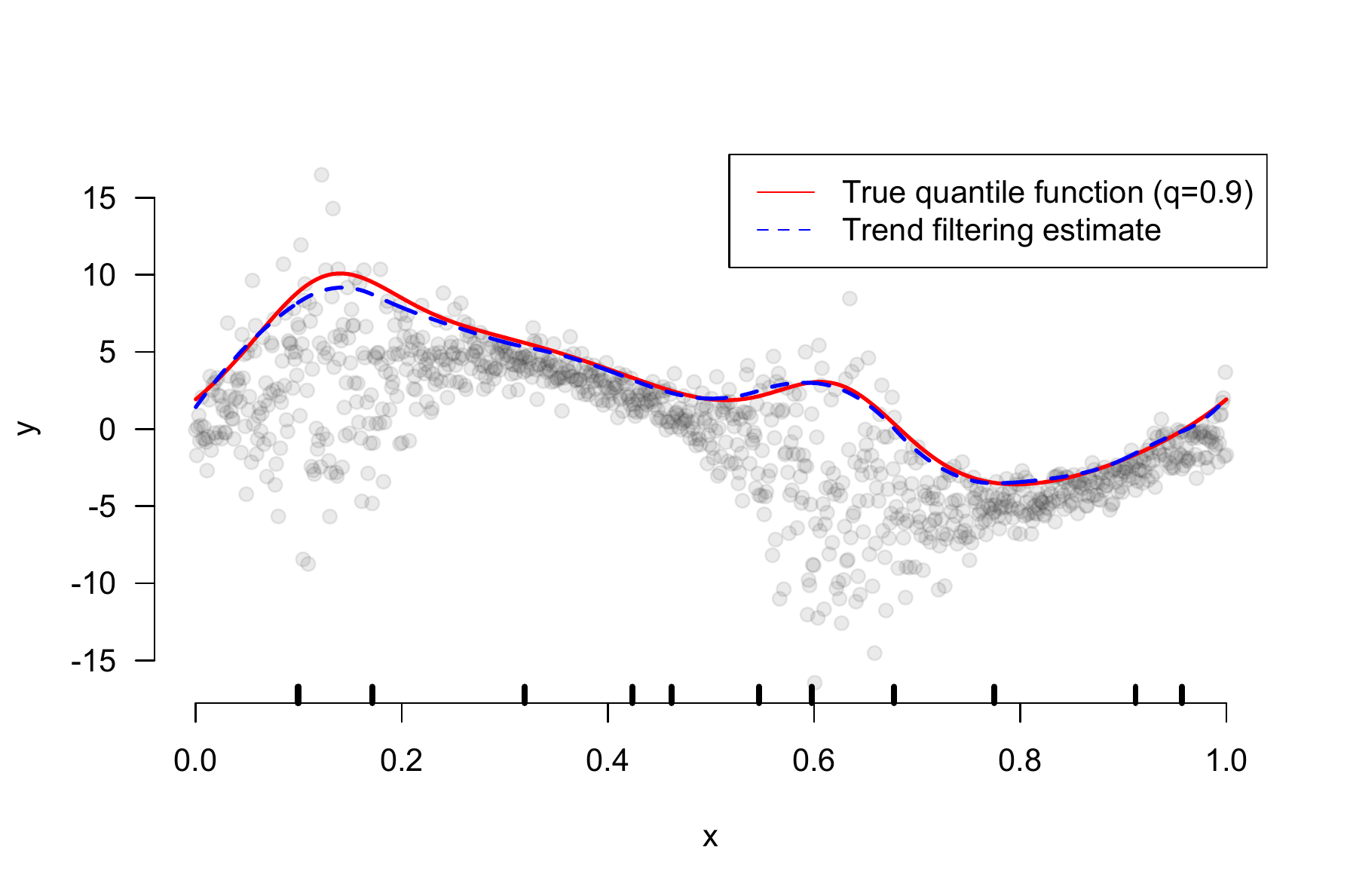}
\caption{\label{fig:qrtf} Quantile-regression trend filtering estimate ($k=2$, $q=0.9$) with the penalty parameter chosen by five-fold cross validation to minimize out-of-sample check loss.  The rug on the $x$ axis shows the adaptively chosen knots in the piece-wise polynomial.}
\end{center}
\end{figure}

To solve the QR trend filtering problem, we use Theorem \ref{thm:varmeanenvelope} to write the objective as
$$
\begin{aligned}
& \underset{\beta \in \R^d}{\text{minimize}}
& & \sum_{i=1}^n \inf_{u_i \geq 0} \left[ \frac{u_i}{2}\left(y_i - \beta_i - \frac{1-2q}{u_i}\right)^2 - \psi(u_i) \right] + \lambda \Vert D^{(t+1)} \beta \Vert_1 \, .
\end{aligned}
$$
where we recall that $\psi(u) = \kappa^2 /( 2u) -1/(2 u^2)$ (see Example \ref{example:quantreg}).

This leads to a simple iterative algorithm where the variational parameter $u$ from Theorem \ref{thm:varmeanenvelope} enters into both the conditional mean and variance of a weighted least squares problem.  Given an estimate $\beta^{(t-1)}$ at step $t-1$ of the algorithm, we form the weights and working responses as
$$
\begin{aligned}
\omega_i^{(t)} & = \frac{\sgn(y_i - \beta_i^{(t-1)})} { y_i - \beta_i^{(t-1)} } \\
z_i^{(t)} & = y_i - (1-2q)/\omega_i^{(t)} \, .
\end{aligned}
$$
Then we update $\beta$ as
$$
\beta^{(t)} = \arg \min_\beta \sum_{i=1}^n \frac{\omega_i^{(t)}}{2} (z_i^{(t)} - \beta_i)^2 + \lambda \Vert D^{(t+1)} \beta \Vert_1 \, .
$$
This substep is an ordinary (weighted) trend-filtering problem and can be solved efficiently using any of several algorithms.  We use the ADMM algorithm described by \citet{ramdas:tibs:2014}, which we have found to be remarkably fast in practice.

Figure \ref{fig:qrtf} shows an example of our QR trend filtering algorithm applied to the following simulated data set where the conditional mean and variance change nonlinearly as a function of $x \in [0,1]$:
$$
\begin{aligned}
y_i & = 5 \sin(2\pi x_i) + e_i \; , \quad  e_i \sim \N(0, \sigma^2(x_i)) \\
\sigma(x_i) & = 0.5 + \exp\{1.5\sin(4\pi x_i)\}\, .
\end{aligned}
$$
We simulated 1000 observations on a regular grid and chose the penalty parameter by five-fold cross validation across a coarse grid $\log_{10}(\lambda) \in \{-1, -0.5, \ldots, 4.5, 5\}$.  The reconstruction quality is excellent, and the algorithm converges rapidly: Figure \ref{thm:varmeanenvelope} shows the results from re-running our algorithm for only 30 steps at the optimal choice of $\lambda$, which took less than half a second on a laptop.

\subsection{Binomial smoothing with the fused double Pareto}

\begin{figure}
\begin{center}
\includegraphics[width=5.6in]{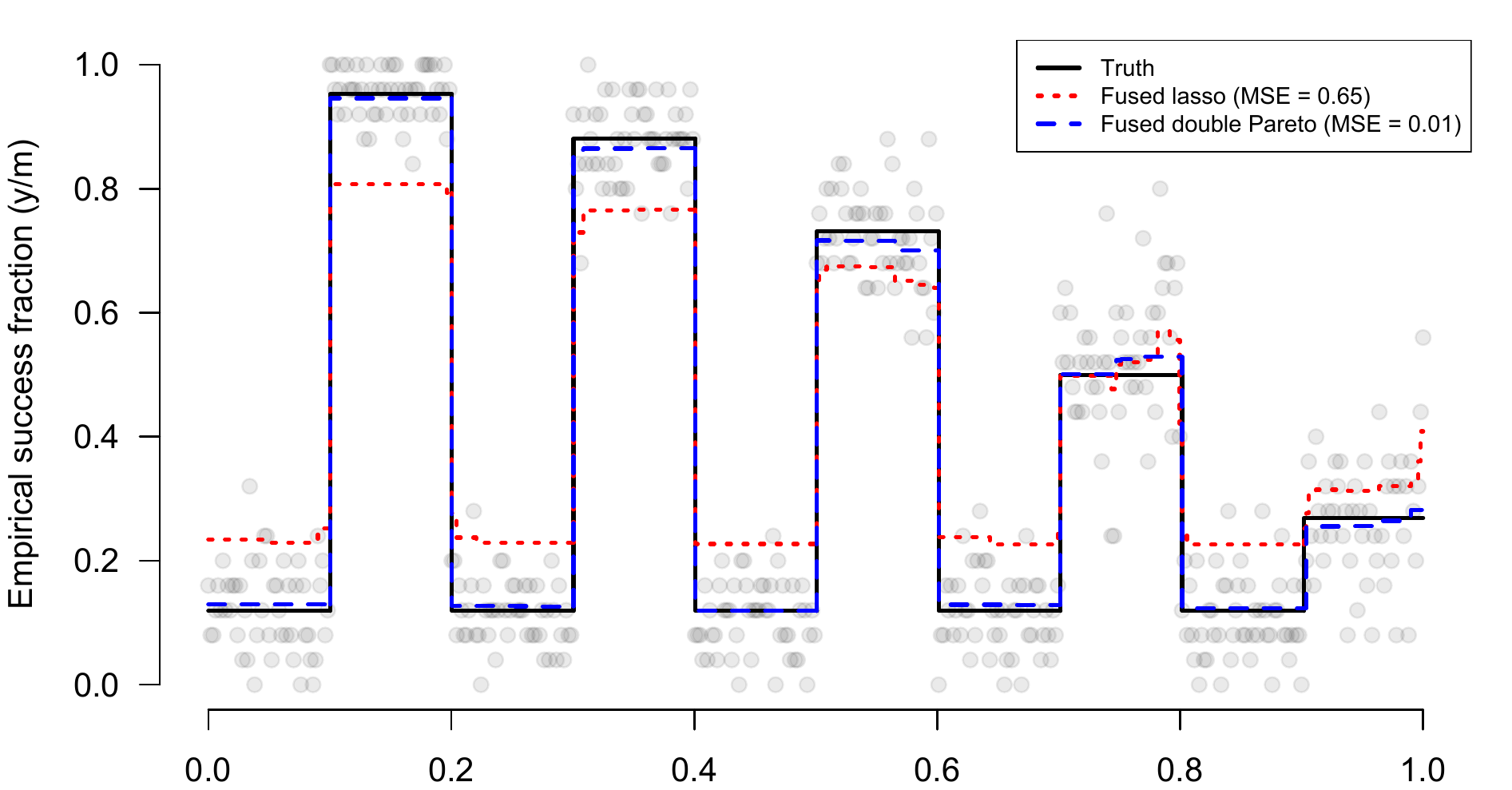}
\caption{\label{fig:fdp} Fused double Pareto versus the fused lasso for a binomial logit problem.  The black line represents the true success fraction as a function of $x$.  The dots represent the empirical success fraction from 25 Bernoulli trials at each point.  The mean-squared error in reconstructing $\beta$ is much smaller for the fused double Pareto than for the fused lasso.}
\end{center}
\end{figure}

Consider observations $y_i$ from the nonlinear  binomial logit model,
$$
y_i \sim \mbox{Binomial}(m_i, w_i) \; , \quad w_i  = 1/\big\{1+e^{-f(x_i)} \big\} \, ,
$$
where $m_i$ is known and the regression function $f(x)$ is assumed to be piecewise constant.  Many authors have considered the use of a fused lasso penalty for recoving $f(x)$.  However, the ``non-diminishing bias'' feature of the $\ell_1$ penalty can often result in over-shrinkage of the estimated parameters \citep[e.g.][]{fan:li:2001}, which in this case would correspond to oversmoothing $\hat{f}(x)$.  To address this potential problem, we implemented a fused double-Pareto model, in which $f(x_i) \equiv \beta_i$ is estimated as the solution to the following optimization problem:
$$
\begin{aligned}
& \underset{\beta \in \R^d}{\text{minimize}}
& & \sum_{i=1}^n \left\{ m_i \log(1 + e^{\beta_i}) - y_i \beta_i ) \right\} + \lambda \sum_{i=2}^n \log(1 + |\beta_{i} - \beta_{i-1}|) \, ,
\end{aligned}
$$
which is defined by combining the negative log likelihood of the logit model and the double-Pareto penalty applied to the first differences of the $\beta$ vector.  By using a concave penalty function, we hope to address the potential problem of over-smoothing.

We applied the results described in Example \ref{example:doublepareto} to re-express this problem as
$$
\begin{aligned}
& \underset{\beta \in \R^d}{\text{minimize}}
& &  l(\beta) + \lambda \sum_{i=2}^n  \inf_{u_i} \left\{ u_i |\beta_i - \beta_{i-1}| + u_i - \lambda \log u_i \right\} \, ,
\end{aligned}
$$
where $l(\beta)$ denotes the loss function.  For fixed $u$, this becomes a fused-lasso problem with a different penalty parameter $u_i$ applied to each first difference.

This leads to the following iterative algorithm.  Given a current estimate $\beta^{(t-1)}$, first we update $u_i$ as 
$$
u_i^{(t)} = \frac{\lambda}{1 + \big| \beta_{i}^{(t-1)} - \beta_{i-1}^{(t-1)} \big|} \, .
$$
Then we update $\beta$ as
$$
\beta^{(t)} = \arg \min_\beta \left\{   l(\beta) +  \sum_{i=2}^n  u_i^{(t)} |\beta_i - \beta_{i-1}| \right\} \, ,
$$
which can be solved using existing methods for the logistic-regression fused lasso.

To illustrate this approach, we simulated data from the binomial logit model in Figure \ref{fig:fdp}.  At 500 evenly spaced points $x_i$ along the unit interval, we simulated 25 Bernoulli trials.  We then fit two solution paths across a grid of $\lambda$ values, one for the fused lasso and one for the fused double-Pareto. For each model we picked $\lambda$ to minimize AIC.  

The solution path of the binomial fused lasso was easy to compute using the methods available in \verb|glmgen| package \citep{glmgen:2014}.  Warm starts were used to improve the speed of convergence.  To compute the solution path for the fused double-Pareto, we initialized the fit at each value of $\lambda$ at the fused-lasso solution for the same $\lambda$.  Our goal was to address the potential problems identified by \citet{mazumder:friedman:hastie:2009} with path algorithms for non-convex problems. 

The results of the comparison are shown in Figure \ref{fig:fdp}.  They show that the fused lasso exhibits a problem with non-diminishing bias (MSE = 0.65), and that the fused double-Pareto successfully addresses the problem (MSE = 0.01).

\section{Discussion}

This paper has presented a framework for representing statistical objective functions in algorithmically convenient ways.  This framework, developed fully in Sections 2--4, connects marginalization with profiling through the notion of hierarchical duality, thereby uniting many previous approaches whose connections have gone unappreciated.

Section \ref{sec:applications} presented three statistical applications that highlight one of the strengths of the framework.  Using our results to derive envelopes requires almost no analytical work.  From there, the updates for the variational parameter are trivial, and the updates for the main parameter can be solved efficiently using existing methods and software.  This modularity offers practitioners the ability to exploit off-the-shelf algorithms for solving ``weighted regression + penalty'' problems efficiently, and can therefore substantially reduce the time and effort that must be invested in exploring novel combinations of loss functions and penalties for a particular data-analysis task.

The immediate motivation for our work was the desire to provide an overarching theory to connect the various special cases of envelope representations studied by \citet{geman:reynolds:1992} and \citet{geman:yang:1995} in image restoration, and more recently by \citet{taddy:2010} and \citet{strawderman:etal:2013} in penalized-likelihood estimation.  In particular, both \citet{taddy:2010} and \citet{strawderman:etal:2013} express surprise and interest in the ``profile Bayesian'' interpretation of the estimators they study.  Yet these authors do not connect their work with the earlier line of thinking on image analysis, or with modern signal-processing algorithms like the proximal gradient method.  One of our goals has been to exploit this connection and generalize it to a broader class of functions that are common in statistics.

We have also sought to answer a much more basic question: under what circumstances does profiling have a sound Bayesian interpretation?  Our results provide at least a partial answer for conditionally exponential and conditionally normal models: for the vast majority of commonly used mixture representations, there is a dual envelope representation, and profiling under the latter is equivalent to marginalizing under the former.  Moreover, the latter is typically much easier to work with, as one never needs to solve an inverse integral equation to identify an appropriate mixing measure.

An important difference between mixture and envelope representations concerns the propriety of the corresponding likelihoods, priors, and posteriors.  If $p(x,\lambda)$ is a proper joint distribution, then $\int p(x, \lambda) d \lambda)$ is proper, but $\sup_{\lambda} p(x,\lambda)$ need not be.  Indeed, we have seen many examples---including quantile regression, the minimax-concave penalty, and logistic regression---in which the function of interest does not correspond to the negative logarithm of a proper probability distribution, but still has an envelope representation in terms of a proper joint distribution.  This is makes the profile approach very useful for handling pseudo-likelioods, improper priors, or likelihoods for discrete parameters that are not proper probability distributions in themselves (as in logistic regression).  There has been some work on representing likelihoods using improper mixing measures \citep[e.g.][]{gramacy:polson:2012}.  But this requires attention to finer points of measure theory, which is unnecessary when using envelope representations.

\appendix

\section{Proofs}

\paragraph{Theorem \ref{thm:dualexponential}.}

The Bernstein--Widder theorem \citep[e.g.][Theorem 12, Chapter IV]{widder:1946} states that a function $f(x)$ is completely monotone if and only if
$$
f(x) = \int_0^{\infty} e^{-\lambda x} d F(\lambda) \, ,
$$
where $F(\lambda)$ is the cumulative distribution function of some non-negative finite Borel measure.  Therefore any density that is a mixture of exponentials must be completely monotone.  Moreover, we also have the following characterization of a completely monotone function in terms of its logarithm \citep[e.g.][Theorem 4.1.5]{bochner:1960}.  Suppose that $\phi(x) \geq 0$.  Then the function $f(x) = e^{-a \phi(x)}$ is completely monotone for every $a > 0$ if and only if $\phi'(x)$ is completely monotone.  This establishes the backward direction.

Moreover, if $e^{-\phi(x)}$ is a mixture of exponentials, then $\phi''(x)$ exists and is nonpositive everywhere.  Thus $\phi(x)$ is concave and has the envelope representation given by the theorem.  This establishes the forward direction.

\paragraph{Theorem \ref{thm:scalemixtureduals}.}

We appeal to the following result on normal scale mixtures from \citet{andrews:mallows:1974}.  Let $f(x)$ be a density function on $\R$.  The composition $g(x) = f(\sqrt{2x})$ is completely monotone if and only if $f$ is a Gaussian scale mixture:
\begin{equation}
\label{eqn:normalscalemixtures}
f(x) = g(x^2/2) = \int_0^{\infty} e^{-\frac{1}{2} \lambda x^2} d F(\lambda) \, .
\end{equation}
This may be seen by applying the Bernstein--Widder theorem to $g(x^2/2)$.

Now let $f(x) = e^{-\phi(x)}$.  We have $f(\sqrt{2x}) = e^{-\theta(x)}$.  Appealing again to Theorem 4.1.5 of \citet{bochner:1960}, $f(\sqrt{2x})$ is completely monotone (and thus a Gaussian scale mixture) if and only if $\theta'(x)$ is completely monotone.

Moreover, if this condition is satisfied, then $\theta''(x) \leq 0$, and $\theta(x)$ is concave. Then
$$
\theta(x) = \inf_{\lambda} \left\{ \lambda x - \theta^\star(\lambda) \right\} \, ,
$$
and therefore
$$
\phi(x) = \inf_{\lambda} \left\{ \frac{\lambda}{2} x^2 - \theta^\star(\lambda) \right\} \, .
$$
This allows us to write $f(x)$ as
$$
f(x) = \sup_{\lambda \geq 0} \left\{ e^{-\frac{\lambda}{2} x^2 + \theta^{\star}(\lambda)}  \right\} = \sup_{\lambda \geq 0} \left\{ \N(x \mid 0, \lambda^{-1}) \ p_V(\lambda) \right \}
$$

The optimal value of $\lambda$ may be computed from the representation $\theta(z) = \inf_{\lambda \geq 0} \{ \lambda z - \theta^\star(\lambda) \}$.  By Lemma \ref{fact:fenchel}, any optimal value of $\lambda$ in this expression satisfies
$$
\hat \lambda(z) \in \partial \theta(z) \, .
$$
As $\phi(x) = \theta(x^2/2)$, we evaluate $\hat \lambda(z)$ at $z=x^2/2$.  In the case of a differentiable $\phi$, this becomes $\phi'(\sqrt{2z})/\sqrt{2z} = \phi'(x)/x$.

\paragraph{Lemma \ref{lemma:gemanyang}.}

Let $\theta^{\star}(\lambda)$ be the dual for $\theta(x) = \frac{1}{2} x^2 - \phi(x)$.  Then after completing the square in $(x-\lambda)$, we have
\begin{eqnarray*}
\theta^{\star}(\lambda) &=& \sup_{x} \left\{ \lambda x - \frac{1}{2}x^2 + \phi(x) \right\} \\
&=& \psi(\lambda) + \frac{1}{2} \lambda^2 \; , \quad \mbox{where} \quad  \psi(\lambda) = \sup_{x} \left\{ - \frac{1}{2}(x-\lambda)^2 + \phi(x) \right\} \, .
\end{eqnarray*}
Because $\theta(x)$ is a closed convex function, $\theta^{\star \star}(x) = \theta(x)$ by Lemma \ref{fact:fenchel}, and so
\begin{eqnarray*}
 \frac{1}{2} x^2 - \phi(x) &=& \sup_{\lambda} \left\{ \lambda x - \theta^{\star}(\lambda) \right\} \\
 &=& -\inf_{\lambda} \left\{ \frac{1}{2} \lambda^2 - \lambda x + \psi(\lambda)  \right\} \, .
\end{eqnarray*}
Therefore
$$
\phi(x) = \inf_{\lambda} \left\{ \frac{1}{2}(x- \lambda)^2 + \psi(\lambda)  \right\} \, ,
$$
proving (A).  To show (B), we apply Lemma \ref{fact:fenchel} to $\theta(x)$ and conclude that a maximizing value of $\lambda$ must satisfy $\hat{\lambda}(x) \in \partial \theta(x) = \{x\} - \partial \phi(x)$, or simply $\hat{\lambda}(x) = x - \phi'(x)$ for differentiable $\phi$.


\paragraph{Theorem \ref{theorem:locationmixture}}

Suppose that condition (4) holds.  Then Lemma 4, together with basic manipulations, are sufficient to verify the existence of the envelope representation as stated.  Now suppose that conditions (1)--(3) hold.  Then Theorem VIII.6.3 of \citet{hirschman:widder:1955}, guarantees that $p(x)$ is a valid solution to the heat equation, or equivalently the Weierstrass transform of a bounded density function.  This proves the existence of a mixture representation.

\paragraph{Theorem \ref{thm:varmeanenvelope}.}

Suppose there exists a $\kappa$ for which $g(x) = f(x) + \kappa x$ is symmetric in $x$, and suppose that $\theta(x) = g(\sqrt{2x})$ is concave on $\R^+$.  We may therefore write $\theta(z)$ in terms of its concave conjugate as $\theta(z) = \inf_{\lambda} \{  \lambda z - \theta^{\star}(\lambda) \}$.  Opening up the definition of $\theta(z)$, we have
$$
\begin{aligned}
f(x) + \kappa x &= \theta(x^2/2) \\
f(x) &= \inf_{\lambda  \geq 0} \left\{ \frac{\lambda}{2} x^2 - \kappa x - \theta^\star(\lambda)  \right\} \\
&=  \inf_{\lambda  \geq 0} \left\{ \frac{\lambda}{2} \left( x - \kappa \lambda^{-1} \right)   - \frac{\kappa^2}{2\lambda}  - \theta^\star(\lambda) \right\} \, ,
\end{aligned}
$$

We use this fact to write $p(x)$ as
$$
\begin{aligned}
p(x) \propto e^{-f(x)} &= \sup_{\lambda \geq 0} \left\{ e^{-\frac{\lambda}{2} \left( x - \kappa \lambda^{-1} \right)} \ e^{\psi(\lambda)} \right\} \\
&= \sup_{\lambda \geq } \left\{ \N(x \mid \kappa \lambda^{-1}, \lambda^{-1}) \ \lambda^{-1/2} e^{\psi(\lambda)} \right\}
\end{aligned}
$$
where $\psi(\lambda) = \frac{\kappa^2}{2\lambda}  + \theta^\star(\lambda)$.  Moreover, the optimal value of $\lambda$ follows from applying Part C of Lemma \ref{fact:fenchel} to the function $\theta(x)$.  If $\phi(x)$ is differentiable, then so is $\theta(z)$, and so
$$
\hat{\lambda}(x) = \theta'(x^2/2) = \frac{f'(x) + \kappa}{x} \, .
$$

\paragraph{Theorem \ref{lemma:lipschitzenvelope}.}

If $f(x) = -\log p(x)$ meets the stated conditions, then $\nabla f$ is Lipschitz continuous with modulus of continuity $L = 1/a$.  By the Cauchy-Schwartz inequality, this implies
$$
\{ \nabla f(x) - \nabla f(y) \}^T (x-y) \leq \frac{L}{2} \enorm{x-y}^2 \, .
$$
Therefore $Lx - \nabla f(x)$ is monotone in $x$, which is equivalent to the function
$$
\theta(x) = \frac{L}{2} \enorm{x}^2 - f(x)
$$
being convex.  Let $\theta^\star(\lambda)$ be the convex conjugate of $\theta(x)$.  Then
$$
\theta(x) = -\inf_{\lambda \in \R^d} \{ \theta^\star(\lambda) - \lambda^T x \} \, ,
$$
and the optimal value of $\lambda$ is, by Lemma \ref{fact:fenchel},
$$
\hat{\lambda}(x) = \nabla \theta(x) = Lx - \nabla f(x) \, .
$$
Equivalently,
$$
f(x) = \frac{L}{2} \enorm{x}^2 - \inf_{\lambda \in \R^d} \left\{ \lambda^T x + \theta^{\star}(\lambda) \right\} \, .
$$
Simple algebra reduces this to
$$
f(x) = \inf_{\lambda \in \R^d} \left\{ \frac{1}{2a} \enorm{x-a\lambda}^2 + \psi(\lambda) \right\} \, ,
$$
with $a=1/L$ and $\psi(\lambda) = \theta^{\star}(\lambda) - \frac{L}{2} \enorm{\lambda}^2$.  Expressing $p(x) = e^{-f(x)}$ in terms of this envelope yields the formula already given.

\begin{small}
\singlespacing
\bibliographystyle{abbrvnat}
\bibliography{masterbib}

\begin{thebibliography}{64}
\providecommand{\natexlab}[1]{#1}
\providecommand{\url}[1]{\texttt{#1}}
\expandafter\ifx\csname urlstyle\endcsname\relax
  \providecommand{\doi}[1]{doi: #1}\else
  \providecommand{\doi}{doi: \begingroup \urlstyle{rm}\Url}\fi

\bibitem[Andrews and Mallows(1974)]{andrews:mallows:1974}
D.~Andrews and C.~Mallows.
\newblock Scale mixtures of normal distributions.
\newblock \emph{Journal of the Royal Statistical Society, Series B},
  36:\penalty0 99--102, 1974.

\bibitem[Armagan(2009)]{armagan:2009}
A.~Armagan.
\newblock Variational bridge regression.
\newblock \emph{Journal of Machine Learning Research W\&CP}, 5\penalty0
  (17--24), 2009.

\bibitem[Armagan et~al.(2012)Armagan, Dunson, and Lee]{dunson:armagan:lee:2010}
A.~Armagan, D.~Dunson, and J.~Lee.
\newblock Generalized double {P}areto shrinkage.
\newblock \emph{Statistica Sinica}, to appear, 2012.

\bibitem[Arnold et~al.(2014)Arnold, Sadhanala, and Tibshirani]{glmgen:2014}
T.~Arnold, V.~Sadhanala, and R.~J. Tibshirani.
\newblock \emph{glmgen: Fast generalized lasso solver}.
\newblock https://github.com/statsmaths/glmgen, 2014.
\newblock {R} package version 0.0.2.

\bibitem[Bae and Mallick(2004)]{bae:mallick:2004}
K.~Bae and B.~Mallick.
\newblock Gene selection using a two-level hierarchical {B}ayesian model.
\newblock \emph{Bioinformatics}, 20\penalty0 (18):\penalty0 3423--30, 2004.

\bibitem[Barndorff-Nielsen(1978)]{barndorffnielsen:1978}
O.~E. Barndorff-Nielsen.
\newblock Hyperbolic distributions and distributions on hyperbolae.
\newblock \emph{Scandinavian Journal of Statistics}, 5\penalty0 (151--7), 1978.

\bibitem[Barndorff-Nielsen et~al.(1982)Barndorff-Nielsen, Kent, and
  Sorensen]{bn:kent:sorensen:1982}
O.~E. Barndorff-Nielsen, J.~Kent, and M.~Sorensen.
\newblock Normal variance-mean mixtures and z distributions.
\newblock \emph{International Statistical Review}, 50:\penalty0 145--59, 1982.

\bibitem[Berger(2006)]{bergerBA2004}
J.~O. Berger.
\newblock The case for objective {B}ayesian analysis.
\newblock \emph{Bayesian Analysis}, 1\penalty0 (3):\penalty0 385--402, 2006.

\bibitem[Bhattacharya et~al.(2012)Bhattacharya, Pati, Pillai, and
  Dunson]{bhattacharya:etal:2012}
A.~Bhattacharya, D.~Pati, N.~S. Pillai, and D.~B. Dunson.
\newblock Bayesian shrinkage.
\newblock http://arxiv.org/abs/1212.6088, 2012.

\bibitem[Bochner(1960)]{bochner:1960}
S.~Bochner.
\newblock \emph{Harmonic Analysis and the Theory of Probability}.
\newblock University of California Press, Berkeley, USA, 1960.

\bibitem[Bogdan et~al.(2011)Bogdan, Chakrabarti, Frommlet, and
  Ghosh]{bogdan:ghosh:2008b}
M.~Bogdan, A.~Chakrabarti, F.~Frommlet, and J.~K. Ghosh.
\newblock Asymptotic {B}ayes-optimality under sparsity of some multiple testing
  procedures.
\newblock \emph{The Annals of Statistics}, 39\penalty0 (3):\penalty0 1551--79,
  2011.

\bibitem[Boyd and Vandenberghe(2004)]{boyd:vandenberghe:2004}
S.~Boyd and L.~Vandenberghe.
\newblock \emph{Convex Optimization}.
\newblock Cambridge University Press, New York, 2004.

\bibitem[Caron and Doucet(2008)]{caron:doucet:2008}
F.~Caron and A.~Doucet.
\newblock Sparse {B}ayesian nonparametric regression.
\newblock In \emph{Proceedings of the 25th International Conference on Machine
  Learning}, pages 88--95. Association for Computing Machinery, Helsinki,
  Finland, 2008.

\bibitem[Carvalho et~al.(2010)Carvalho, Polson, and
  Scott]{Carvalho:Polson:Scott:2008a}
C.~M. Carvalho, N.~G. Polson, and J.~G. Scott.
\newblock The horseshoe estimator for sparse signals.
\newblock \emph{Biometrika}, 97\penalty0 (2):\penalty0 465--80, 2010.

\bibitem[Combettes and Pesquet(2011)]{combettes:pesquet:2011}
P.~Combettes and J.~Pesquet.
\newblock Proximal splitting methods in signal processing.
\newblock In \emph{Fixed-Point Algorithms for Inverse Problems in Science and
  Engineering}, Optimization and its Applications. Springer, 2011.

\bibitem[Dasgupta(1994)]{dasgupta:1994}
A.~Dasgupta.
\newblock Distributions which are {G}aussian convolutions.
\newblock In S.~S. Gupta and J.~O. Berger, editors, \emph{Statistical Decision
  Theory and Related Topics}, volume~V, pages 391--400. Springer--Verlag, 1994.

\bibitem[Datta and Ghosh(2013)]{datta:ghosh:2013}
J.~Datta and J.~Ghosh.
\newblock Asymptotic properties of {B}ayes risk for the horseshoe prior.
\newblock \emph{Bayesian Analysis}, 8\penalty0 (1):\penalty0 111--32, 2013.

\bibitem[Dempster et~al.(1977)Dempster, Laird, and
  Rubin]{dempster:laird:rubin:1977}
A.~Dempster, N.~Laird, and D.~Rubin.
\newblock Maximum likelihood from incomplete data via the {EM} algorithm (with
  discussion).
\newblock \emph{Journal of the Royal Statistical Society (Series B)},
  39\penalty0 (1):\penalty0 1--38, 1977.

\bibitem[Efron and Morris(1972)]{efron:morris:1972}
B.~Efron and C.~Morris.
\newblock Limiting the risk of {B}ayes and empirical {B}ayes---part ii: the
  empirical {B}ayes case.
\newblock \emph{Journal of the American Statistical Association}, 67\penalty0
  (337):\penalty0 130--9, 1972.

\bibitem[Efron et~al.(2004)Efron, Hastie, Johnstone, and
  Tibshirani]{efron:LARS:2004}
B.~Efron, T.~Hastie, I.~Johnstone, and R.~Tibshirani.
\newblock Least angle regression.
\newblock \emph{The Annals of Statistics}, 32\penalty0 (2):\penalty0 407--99,
  2004.

\bibitem[Fan and Li(2001)]{fan:li:2001}
J.~Fan and R.~Li.
\newblock Variable selection via nonconcave penalized likelihood and its oracle
  properties.
\newblock \emph{Journal of the American Statistical Association}, 96\penalty0
  (456):\penalty0 1348--60, 2001.

\bibitem[Figueiredo(2003)]{figueiredo:2003}
M.~Figueiredo.
\newblock Adaptive sparseness for supervised learning.
\newblock \emph{IEEE Transactions on Pattern Analysis and Machine
  Intelligence}, 25\penalty0 (9):\penalty0 1150--9, 2003.

\bibitem[Figueiredo and Nowak(2003)]{fig:nowak:2003}
M.~Figueiredo and R.~Nowak.
\newblock An {EM} algorithm for wavelet-based image restoration.
\newblock \emph{IEEE Transactions on Image Processing}, 12:\penalty0 906--16,
  2003.

\bibitem[Friedman et~al.(2010)Friedman, Hastie, and
  Tibshirani]{fried:hastie:tibs:2010}
J.~H. Friedman, T.~Hastie, and R.~Tibshirani.
\newblock Regularization paths for generalized linear models via coordinate
  descent.
\newblock \emph{Journal of Statistical Software}, 2010.

\bibitem[Geman and Reynolds(1992)]{geman:reynolds:1992}
D.~Geman and G.~Reynolds.
\newblock Constrained restoration and the recovery of discontinuities.
\newblock \emph{IEEE Transactions on Pattern Analysis and Machine
  Intelligence}, 14\penalty0 (3):\penalty0 367--83, 1992.

\bibitem[Geman and Yang(1995)]{geman:yang:1995}
D.~Geman and C.~Yang.
\newblock Nonlinear image recovery with half-quadratic regularization.
\newblock \emph{IEEE Transactions on Image Processing}, 4\penalty0
  (7):\penalty0 932--46, 1995.

\bibitem[Gramacy and Polson(2012)]{gramacy:polson:2012}
R.~B. Gramacy and N.~G. Polson.
\newblock Simulation-based regularized logistic regression.
\newblock \emph{Bayesian Analysis}, 7\penalty0 (3):\penalty0 567--90, 2012.

\bibitem[Griffin and Brown(2005)]{griffin:brown:2005}
J.~Griffin and P.~Brown.
\newblock Alternative prior distributions for variable selection with very many
  more variables than observations.
\newblock Technical report, University of Warwick, 2005.

\bibitem[Griffin and Brown(2010)]{griffin:brown:2010}
J.~Griffin and P.~Brown.
\newblock Inference with normal-gamma prior distributions in regression
  problems.
\newblock \emph{Bayesian Analysis}, 5\penalty0 (1):\penalty0 171--88, 2010.

\bibitem[Hahn and Carvalho(2013)]{hahn:carvalho:2013}
P.~Hahn and C.~M. Carvalho.
\newblock Decoupling shrinkage and selection in bayesian linear models.
\newblock Technical report, University of Chicago Booth School of Business,
  2013.
\newblock URL
  \url{http://faculty.chicagobooth.edu/richard.hahn/HahnCarvalhoDSS2013.pdf}.

\bibitem[Hans(2009)]{hans:2008}
C.~M. Hans.
\newblock {B}ayesian lasso regression.
\newblock \emph{Biometrika}, 96\penalty0 (4):\penalty0 835--45, 2009.

\bibitem[Hans(2011)]{hans:2011}
C.~M. Hans.
\newblock Elastic net regression modeling with the orthant normal prior.
\newblock \emph{Journal of the American Statistical Association}, to appear,
  2011.

\bibitem[Hirschman and Widder(1955)]{hirschman:widder:1955}
I.~Hirschman and D.~V. Widder.
\newblock \emph{The Convolution Transform}.
\newblock Princeton University Press, Princeton, New Jersey, 1955.

\bibitem[Jaakkola and Jordan(2000)]{jaakkola:jordan:2000}
T.~Jaakkola and M.~I. Jordan.
\newblock Bayesian parameter estimation via variational methods.
\newblock \emph{Statistics and Computing}, 10\penalty0 (25--37), 2000.

\bibitem[Johnson(2013)]{johnson:2013}
N.~Johnson.
\newblock A dynamic programming algorithm for the fused lasso and $l$-0
  segmentation.
\newblock \emph{Journal of Computational and Graphical Statistics}, 22\penalty0
  (2):\penalty0 246--60, 2013.

\bibitem[Kim et~al.(2007)Kim, Koh, Lustig, and Boyd]{boyd:l1:2007}
S.-J. Kim, K.~Koh, M.~Lustig, and S.~Boyd.
\newblock An interior-point method for large-scale regularized least squares.
\newblock \emph{IEEE Journal of Selected Topics in Signal Processing},
  1\penalty0 (4):\penalty0 606--17, 2007.

\bibitem[Kim et~al.(2009)Kim, Koh, Boyd, and Gorinevsky]{kim:boyd:etal:2009}
S.-J. Kim, K.~Koh, S.~Boyd, and D.~Gorinevsky.
\newblock $\ell^1$ trend filtering.
\newblock \emph{SIAM Reviews}, 51:\penalty0 339--60, 2009.

\bibitem[Koenker(2005)]{koenker:2005}
R.~Koenker.
\newblock \emph{Quantile {R}egression}.
\newblock Cambridge University Press, New York, USA, 2005.

\bibitem[Leeb and P\"otscher(2008)]{leeb:potscher:2005}
H.~Leeb and B.~P\"otscher.
\newblock Sparse estimators and the oracle property, or the return of {H}odges'
  estimator.
\newblock \emph{Journal of Econometrics}, 142\penalty0 (1):\penalty0 201--11,
  2008.

\bibitem[Li et~al.(2010)Li, Xi, and Lin]{li:xi:lin:2010}
Q.~Li, R.~Xi, and N.~Lin.
\newblock Bayesian regularized quantile regression.
\newblock \emph{Bayesian Analysis}, 5\penalty0 (3):\penalty0 533--56, 2010.

\bibitem[Mazumder et~al.(2011)Mazumder, Friedman, and
  Hastie]{mazumder:friedman:hastie:2009}
R.~Mazumder, J.~Friedman, and T.~Hastie.
\newblock Sparsenet: coordinate descent with non-convex penalties.
\newblock \emph{Journal of the American Statistical Association}, 106\penalty0
  (495):\penalty0 1125--38, 2011.

\bibitem[Muller et~al.(2006)Muller, Parmigiani, and Rice]{muller:etal:2006}
P.~Muller, G.~Parmigiani, and K.~Rice.
\newblock {FDR} and {B}ayesian multiple comparisons rules.
\newblock In \emph{Proceedings of the 8th Valencia World Meeting on Bayesian
  Statistics}. Oxford University Press, 2006.

\bibitem[O'Hagan(1976)]{ohagan:1976}
A.~O'Hagan.
\newblock On posterior joint and marginal modes.
\newblock \emph{Biometrika}, 63\penalty0 (2):\penalty0 329--33, 1976.

\bibitem[Palmer et~al.(2006)Palmer, Wipf, Kreutz-Delgado, and
  Rao]{palmer:etal:2006}
A.~Palmer, D.~Wipf, K.~Kreutz-Delgado, and B.~Rao.
\newblock Variational {EM} algorithms for non-{G}aussian latent variable
  models.
\newblock In \emph{Advances in Neural Information Processing Systems 18}, 2006.

\bibitem[Park and Casella(2008)]{park:casella:2008}
T.~Park and G.~Casella.
\newblock The {B}ayesian lasso.
\newblock \emph{Journal of the American Statistical Association}, 103\penalty0
  (482):\penalty0 681--6, 2008.

\bibitem[Polson and Scott(2012)]{Polson:Scott:2010b}
N.~G. Polson and J.~G. Scott.
\newblock Local shrinkage rules, {L}\'evy processes, and regularized
  regression.
\newblock \emph{Journal of the Royal Statistical Society (Series B)},
  74\penalty0 (2):\penalty0 287--311, 2012.

\bibitem[Polson and Scott(2013)]{Polson:Scott:2011a}
N.~G. Polson and J.~G. Scott.
\newblock Data augmentation for non-{G}aussian regression models using
  variance-mean mixtures.
\newblock \emph{Biometrika}, 100\penalty0 (2):\penalty0 459--71, 2013.

\bibitem[Polson and Scott(2011)]{polson:stevescott:2011}
N.~G. Polson and S.~Scott.
\newblock Data augmentation for support vector machines (with discussion).
\newblock \emph{Bayesian Analysis}, 6\penalty0 (1):\penalty0 1--24, 2011.

\bibitem[Polson et~al.(2013)Polson, Scott, and
  Windle]{polson:scott:windle:2012a}
N.~G. Polson, J.~G. Scott, and J.~Windle.
\newblock Bayesian inference for logistic models using {P}olya-{G}amma latent
  variables.
\newblock \emph{Journal of the American Statistical Association}, 108\penalty0
  (504):\penalty0 1339--49, 2013.

\bibitem[Polson et~al.(2014)Polson, Scott, and Windle]{Polson:Scott:2011d}
N.~G. Polson, J.~G. Scott, and J.~Windle.
\newblock The {B}ayesian bridge.
\newblock \emph{Journal of the Royal Statistical Society (Series B)}, 2014.
\newblock to appear (DOI: 10.1111/rssb.12042).

\bibitem[Ramdas and Tibshirani(2014)]{ramdas:tibs:2014}
A.~Ramdas and R.~J. Tibshirani.
\newblock Fast and flexible {ADMM} algorithms for trend filtering.
\newblock Technical report, Carnegie Mellon University,
  http://www.stat.cmu.edu/~ryantibs/papers/fasttf.pdf, 2014.

\bibitem[Rockafellar and Wets(1998)]{rockafellar:wets:1998}
R.~T. Rockafellar and R.~J.-B. Wets.
\newblock \emph{Variational Analysis}.
\newblock Springer, 1998.

\bibitem[Scott and Berger(2006)]{scottberger06}
J.~G. Scott and J.~O. Berger.
\newblock An exploration of aspects of {B}ayesian multiple testing.
\newblock \emph{Journal of Statistical Planning and Inference}, 136\penalty0
  (7):\penalty0 2144--2162, 2006.

\bibitem[Strawderman et~al.(2013)Strawderman, Wells, and
  Schifano]{strawderman:etal:2013}
R.~Strawderman, M.~T. Wells, and E.~D. Schifano.
\newblock Hierarchical {B}ayes, maximum \textit{a posteriori} estimators, and
  minimax concave penalized likelihood estimation.
\newblock \emph{Electronic Journal of Statistics}, 7:\penalty0 973--90, 2013.

\bibitem[Taddy(2013)]{taddy:2010}
M.~Taddy.
\newblock Multinomial inverse regression for text analysis.
\newblock \emph{Journal of the American Statistical Association}, 108\penalty0
  (503):\penalty0 755--70, 2013.
\newblock arXiv:1012.2098v2.

\bibitem[Tibshirani(2014)]{tibs:2014a}
R.~Tibshirani.
\newblock Adaptive piecewise polynomial estimation via trend filtering.
\newblock \emph{Annals of Statistics}, 42\penalty0 (1):\penalty0 285--323,
  2014.

\bibitem[Tibshirani et~al.(2005)Tibshirani, Saunders, Rosset, Zhu, and
  Knight]{tibs:fusedlasso:2005}
R.~Tibshirani, M.~Saunders, S.~Rosset, J.~Zhu, and K.~Knight.
\newblock Sparsity and smoothness via the fused lasso.
\newblock \emph{Journal of the Royal Statistical Society (Series B)},
  67:\penalty0 91--108, 2005.

\bibitem[Tipping(2001)]{tipping:2001}
M.~Tipping.
\newblock Sparse {B}ayesian learning and the relevance vector machine.
\newblock \emph{Journal of Machine Learning Research}, 1:\penalty0 211--44,
  2001.

\bibitem[Wainwright and Jordan(2008)]{wainwright:jordan:2008}
M.~J. Wainwright and M.~I. Jordan.
\newblock Graphical models, exponential families, and variational inference.
\newblock \emph{Foundations and Trends in Machine Learning}, 1\penalty0
  (1--2):\penalty0 1--305, 2008.

\bibitem[Widder(1946)]{widder:1946}
D.~Widder.
\newblock \emph{The Laplace Transform}.
\newblock Princeton University Press, 1946.

\bibitem[Zhang(2010)]{zhang:2010}
C.-H. Zhang.
\newblock Nearly unbiased variable selection under minimax concave penalty.
\newblock \emph{Annals of Statistics}, 38:\penalty0 894--942, 2010.

\bibitem[Zhang et~al.(2013)Zhang, Qian, Chen, and Zhang]{zhang:etal:2013}
S.~Zhang, H.~Qian, W.~Chen, and Z.~Zhang.
\newblock A concave conjugate approach for nonconvex penalized regression with
  the {MCP} penalty.
\newblock In \emph{Twenty-Seventh AAAI Conference on Artificial Intelligence},
  pages 1027--33, 2013.

\bibitem[Zhou et~al.(2010)Zhou, Lange, and Suchard]{zhou:lange:suchard:2010}
H.~Zhou, K.~Lange, and M.~Suchard.
\newblock Graphics processing units and high-dimensional optimization.
\newblock \emph{Statistical Science}, 25\penalty0 (3):\penalty0 311--24, 2010.

\bibitem[Zou and Li(2008)]{zou:li:2008}
H.~Zou and R.~Li.
\newblock One-step sparse estimates in nonconcave penalized likelihood models.
\newblock \emph{Annals of Statistics}, 36\penalty0 (4):\penalty0 1509--33,
  2008.

\end{thebibliography}

\end{small}

\end{document}